\documentclass[twocolumn,pra,twocolumn,superscriptaddress]{revtex4}
\usepackage{color}
\usepackage{amsmath}
\usepackage{amssymb}
\usepackage{graphicx}
\usepackage{tikz}

\usepackage{float}
\usepackage{bbm}

\usepackage{epsfig} 
\usepackage{verbatim}
\usepackage{array}
\usepackage{setspace}

\usepackage[unicode=true,
 bookmarks=true,bookmarksopen=false,
 breaklinks=false,pdfborder={0 0 1},backref=false,colorlinks=true]
 {hyperref}
\hypersetup{
 linkcolor=magenta, urlcolor=blue, citecolor=blue, pdfstartview={FitH}, hyperfootnotes=false, unicode=true}

\makeatletter
\@ifundefined{textcolor}{}
{%
 \definecolor{BLACK}{gray}{0}
 \definecolor{WHITE}{gray}{1}
 \definecolor{RED}{rgb}{1,0,0}
 \definecolor{GREEN}{rgb}{0,1,0}
 \definecolor{BLUE}{rgb}{0,0,1}
 \definecolor{CYAN}{cmyk}{1,0,0,0}
 \definecolor{MAGENTA}{cmyk}{0,1,0,0}
 \definecolor{YELLOW}{cmyk}{0,0,1,0}
}

\usepackage{amsfonts}\usepackage{tabularx}\usepackage{dcolumn}\usepackage{bm}\usepackage{graphicx}\usepackage{epstopdf}

\hypersetup{urlcolor=blue}

\makeatother

\newcolumntype{C}[1]{>{\centering\arraybackslash$}p{#1}<{$}}

\begin{document}

\title{Theory on electron-phonon spin dehphasing in GaAs  multi-electron double quantum dots}

\author{Guanjie He}
\affiliation{Department of Physics, City University of Hong Kong, Tat Chee Avenue, Kowloon, Hong Kong SAR, China, and City University of Hong Kong Shenzhen Research Institute, Shenzhen, Guangdong 518057, China}
\author{Guo Xuan Chan}
\affiliation{Department of Physics, City University of Hong Kong, Tat Chee Avenue, Kowloon, Hong Kong SAR, China, and City University of Hong Kong Shenzhen Research Institute, Shenzhen, Guangdong 518057, China}
\author{Xin Wang}
\email{x.wang@cityu.edu.hk}
\affiliation{Department of Physics, City University of Hong Kong, Tat Chee Avenue, Kowloon, Hong Kong SAR, China, and City University of Hong Kong Shenzhen Research Institute, Shenzhen, Guangdong 518057, China}
\date{\today}

\begin{abstract}
Recent studies reveal that a double-quantum-dot system hosting more than two electrons may be superior in certain aspects as compared to the traditional case in which only two electrons are confined (a singlet-triplet qubit). We study the electron-phonon dephasing occurring in a GaAs multi-electron double-quantum-dot system, in a biased case in which the singlet state is hybridized, as well as in an unbiased case in which the hybridization is absent. We have found that while the electron-phonon dephasing rate increases with the number of electrons confined in the unbiased case, this does not hold in the biased case. We define a merit figure as a ratio between the exchange energy and the dephasing rate, and have shown that in experimentally relevant range of the exchange energy, the merit figure actually increases with the number of electrons in the biased case. Our results show that the multi-electron quantum-dot system has another advantage in mitigating the effect of electron-phonon dephasing, which is previously under-appreciated in the literature.


\end{abstract}

\maketitle

\section{introduction}

Semiconductor quantum-dot spin qubits as platforms for the physical realization of quantum computation, have attracted extensive research interests due to their promises of tunability, scalability and high-fidelity gate operations \cite{PhysRevA.57.120,10.1093/nsr/nwy153,PhysRevA.98.032334,PhysRevB.59.2070,PhysRevResearch.2.012062,PhysRevLett.105.246804,10.1038/nature23022,PhysRevB.96.201307,PhysRevB.82.045311,PRXQuantum.2.040306,10.1038/s41534-021-00449-4,PhysRevLett.122.217702,PhysRevB.75.081303,PhysRevB.102.035427,PhysRevB.84.235309,PhysRevB.68.115306,PhysRevB.85.075416,10.1002/qute.201900072,PhysRevB.103.L161409,szabo1982modern,maune2012coherent,PhysRevX.11.041025,doi:10.1126/science.aao5965,10.1038/s41467-020-17865-3,del1999electronic}. While
each quantum dot typically hosts no more than two electrons in traditional spin qubits,
recent researches reveal that the multi-electron qubits, in which certain dot is allowed to host more than two electrons,  may be  advantageous in some aspects 
\cite{10.1038/s41467-019-09194-x,mills2021two,PhysRevB.105.075430,10.1063/1.4869108,PhysRevLett.114.226803,10.1021/acs.nanolett.9b02149,Leon.21,PhysRevB.104.L081409,PhysRevB.90.195424,PhysRevLett.127.086802,PhysRevX.8.011045,PhysRevLett.119.227701,science.278.5344.1788,PhysRevLett.112.026801,PhysRevB.91.155425,PhysRevB.88.161408,PhysRevX.10.041010,PhysRevB.97.245301}. For example, a multi-electron quantum dot may serve as a mediator for fast spin exchange \cite{10.1038/s41467-019-09194-x} or a tunable coupling between nearby dots \cite{PhysRevLett.114.226803}.  
Moreover, it has been shown that multi-electron quantum-dot devices may be more resilient to noises than traditional ones due to the screening effect by core electrons
\cite{PhysRevLett.112.026801,PhysRevB.91.155425,PhysRevB.88.161408}.

Experiments show that in certain asymmetric multi-electron triple-quantum-dot system, the dependence of the exchange energy on the absolute value of detuning can be non-monotonic, implying the existence of a sweet spot \cite{PhysRevX.8.011045}. It has also been observed in a similar system that the sign of the exchange energy may reverse, removing a long-standing constraint for the construction of dynamically corrected exchange gates \cite{PhysRevLett.119.227701}.
On the theory side, calculations based on the Configuration Interaction (CI) techniques on few-electron multi-quantum-dot systems have demonstrated negative exchange interactions and their implication on robust quantum control \cite{PhysRevB.97.245301,chan2022microscopic,chan2022robust}. 
Other studies on these systems have unveiled their potentials for tunable couplings \cite{PhysRevLett.114.226803}, robust quantum gates \cite{PhysRevB.88.161408}, as well as other interesting properties  \cite{PhysRevX.10.041010}. These results have shown the promises of multi-electron quantum-dot systems in achieving noise-resilient quantum information processing.


Various environmental noises and ways to combat them have been extensively studied in conventional two-electron singlet-triplet qubits \cite{zhao2018toward,kornich2018phonon,PhysRevB.89.085410,PhysRevB.83.165322,PhysRevB.65.245213,PhysRevLett.110.146804,PhysRevLett.96.100501,yoneda2018quantum,doi:10.1126/science.1217692,PhysRevLett.116.110402,huang2021dephasing,PhysRevB.76.035315,PhysRevB.86.035302,Madzik.20,PhysRevB.100.125430,yoneda2021coherent,gaudreau2012coherent,roszak2015decoherence,10.1038/ncomms11170,10.1038/nphys627,doi:10.1126/science.1159221,mavadia2017prediction,nakajima2018coherent}. Among these noises, the electron-phonon dephasing is an important channel  leading to decoherence \cite{zhao2018toward,PhysRevB.89.085410,kornich2018phonon,PhysRevB.83.165322,PhysRevB.65.245213}. Phonon couplings that contribute to decohence in GaAs include the deformation potential interaction, the polar optical interaction, and the piezoelectric interaction \cite{PhysRevB.86.035302}. In double quantum dots (DQD) hosting two electrons, it has been shown that the deformation potential and  piezoelectric interaction play major roles in the electron-phonon dephasing, and all channels of  phonon couplings reduce as the dot distance increases \cite{PhysRevB.83.165322,PhysRevB.89.085410}. It is an interesting open question how the behavior of the  electron-phonon dephasing may change as the number of electrons in the DQD is increased \cite{chan2022robust}.

In this paper, we investigate the electron-phonon dephasing in a GaAs multi-electron DQD system, in which the electron configurations are more complicated than the case with only two electrons. Nevertheless, we have defined a merit figure as a ratio between the exchange energy and the dephasing rate, and have shown that in experimentally relevant range of the exchange energy, the merit figure actually increases with the number of electrons. These results suggest that the multi-electron quantum-dot system have advantages in reducing noises stemming from the electron-phonon interaction, which is previously under-appreciated in the literature.


The remainder of the paper is organized as follows. In Sec.~\ref{sec:model} we present our model of the multi-electron DQD system, and methods to solve the electron-phonon interaction problem. Sec.~\ref{sec:result} shows the results on the dephasing rates, exchange energies and the merit figures in different cases. In the end we conclude in Sec.~\ref{sec:conclusion}.

\section{Model and Methods}
\label{sec:model}
\subsection{Hamiltonian}

\begin{figure}
(a)\includegraphics[scale=0.42]{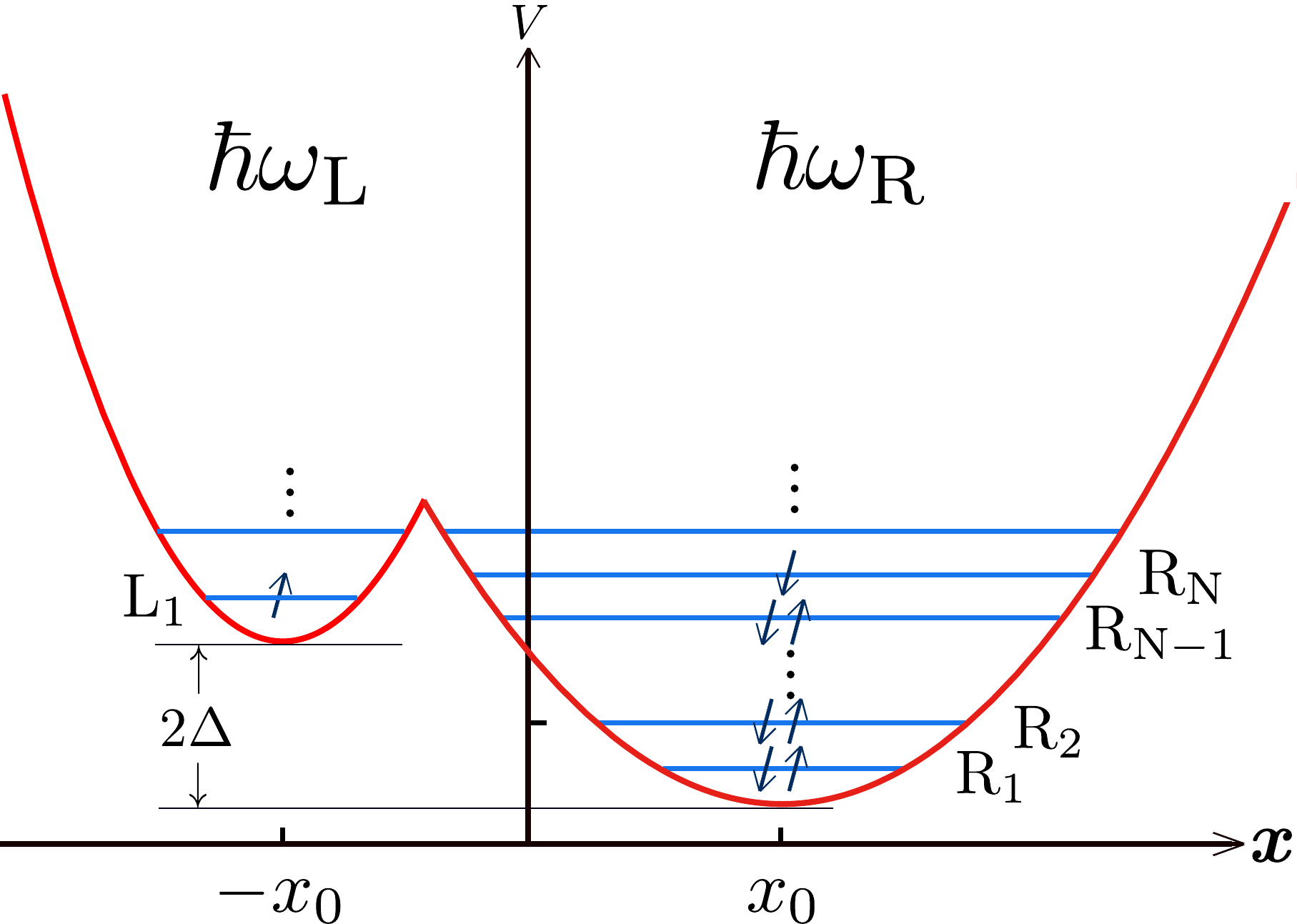}
\ \ \\
\ \ \\
(b)\includegraphics[scale=0.40]{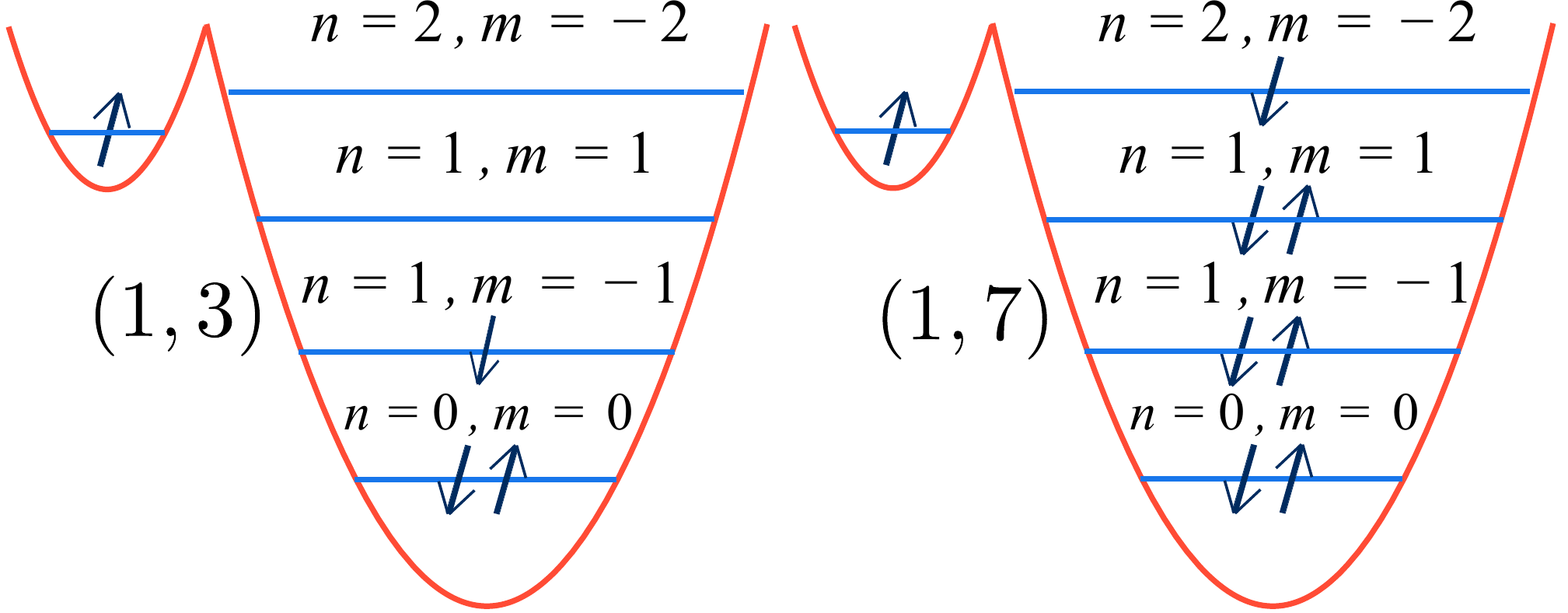}
\caption{(a) Schematic illustration of a double-quantum-dot system hosting $2N$ electrons. One electron occupies the $\mathrm{L}_1$ orbital of the left dot, while $2N-1$ electrons occupy the $\textrm{R}_{1}$ through $\textrm{R}_{N}$ orbitals of the right dot.  (b) Cases with electron configuration $(1,3)$ and $(1,7)$ considered in this paper. Here, $n$ is the principle quantum number of the relevant Fock-Darwin state, and $m$ the magnetic quantum number.}\label{fig:quantumdotmodel}
\end{figure}

We consider an asymmetric double-quantum-dot system where the right dot (R) is larger than the left dot (L), and the distance between the center of the two dots is $2x_0$. We keep the lowest $N$ orbitals in the right dot and label the energy levels from the ground state to the $N$th orbital by $\mathrm{R}_1$ through $\mathrm{R}_N$ as shown in Fig.~\ref{fig:quantumdotmodel}. We assume that the system hosts a total of $2N$ electrons with one electron in the left dot (L) and $2N-1$ electrons in the right dot (R). The Hamiltonian of the system can be written as
\begin{equation}
H=\sum_{j}^{2N}h_{j}+\sum_{j,k}^{2N}\frac{e^{2}}{\epsilon \left | \mathbf{\mathrm{R}_{j}-\mathrm{R}_{k}} \right |},
\end{equation}
where
\begin{equation}
h_{i}=\frac{(-i\hbar\nabla_{j}+e\mathbf{A}/c )^{2}}{2m^{\ast }}+V(\mathbf{r})+g^{\ast }\mu _{B}\mathbf{B}\cdot \mathbf{S}.
\end{equation}
The confinement potential in the $xy$ plane is
\begin{equation}
V(\mathbf{r})=\frac{1}{2}m^{\ast }\mathrm{Min}[\omega _{\textrm{R}}^{2}(\mathbf{r}-\mathbf{r}_{0})^{2}-\Delta,\omega _{\textrm{L}}^{2}(\mathbf{r}+\mathbf{r}_{0})^{2}+\Delta],
\end{equation}
where $\omega _{\textrm{L}}$ ($\omega _{\textrm{R}}$) is the confinement strength in dot L (R),
$\mathbf{r}=(x,y)$, $\mathbf{r}_{0}=(x_{0},0)$, $x_{0}\in\left[40 \mathrm{nm} ,70 \mathrm{nm}\right]$ is half the distance between the center of the two dots, $m^{\ast }=0.067m_{e}$ is the effective mass of the electron, and $\Delta$ is the detuning, as indicated in Fig.~\ref{fig:quantumdotmodel}. In our model, $\hbar\omega _{\textrm{L}}>\hbar\omega _{\textrm{R}}$ because right dot contains more electrons. We therefore fix $\hbar \omega _{\textrm{L}}=2.838\textrm{meV}$ and vary $\hbar \omega _{\textrm{R}}$ between $\hbar \omega _{\textrm{L}}/4$ and $\hbar \omega _{\textrm{L}}/2$. We consider  the case of zero perpendicular magnetic field $B$, but our main conclusion remains for finite $B$.


The system is initialized, at an appropriate value of $\Delta$, in a state with electron occupancy $(1,2N-1)$ where the first (second) entry indicates the occupancy in the L (R) dot. This is called an unbiased case. In the unbiased case, the lowest energy levels are a singlet state, represented by $|\mathrm{S}(1,2N-1)\rangle$, and a non-magnetic triplet state, represented by $|\mathrm{T}(1,2N-1)\rangle$. As $\Delta$ is changed, the system enters a biased case in which $|\mathrm{S}(1,2N-1)\rangle$ hybridizes with $|\mathrm{S}(0,2N)\rangle$ and $|\mathrm{S}(2,2N-2)\rangle$. A key quantity is the exchange energy,
\begin{equation}
J=E_{|\textrm{T}\rangle}-E_{|\textrm{S}\rangle},\label{eq:defJ}
\end{equation}
which we calculate under the Hund-Mulliken approximation \cite{PhysRevB.89.085410,PhysRevB.91.035301}.


\subsection{Singlet and triplet in multi-electron double quantum dot}

In the unbiased case, the singlet state is $|\mathrm{S}(1,2N-1)\rangle$ and the triplet state $|\mathrm{T}(1,2N-1)\rangle$ can be written as

\begin{eqnarray}
	\label{eq:singlet2n1}
	|\mathrm{S}(1,2N-1)\rangle=\frac{1}{\sqrt{2(1+\mathcal{I}_{N,\mathrm{S}})}}(|\uparrow_{\textrm{L}_{1}}\downarrow_{\textrm{R}_{N}}...\uparrow_{\textrm{R}_{1}}\downarrow_{\textrm{R}_{1}}\rangle+\notag\\
	|\uparrow_{\textrm{R}_{N}}\downarrow_{\textrm{L}_{1}}...\uparrow_{\textrm{R}_{1}}\downarrow_{\textrm{R}_{1}}\rangle),\notag\\
\end{eqnarray}
\begin{eqnarray}
	\label{eq:triplet2n1}
	|\mathrm{T}(1,2N-1)\rangle=\frac{1}{\sqrt{2(1-\mathcal{I}_{N,\mathrm{T}})}}(|\uparrow_{\textrm{L}_{1}}\downarrow_{\textrm{R}_{N}}...\uparrow_{\textrm{R}_{1}}\downarrow_{\textrm{R}_{1}}\rangle-\notag\\
	|\uparrow_{\textrm{R}_{N}}\downarrow_{\textrm{L}_{1}}...\uparrow_{\textrm{R}_{1}}\downarrow_{\textrm{R}_{1}}\rangle),\notag\\
\end{eqnarray}
where $\textrm{L}_1$ and $\textrm{R}_{i}$ $(i=1\ldots N)$ label the orbital states occupied by electrons in the left and right dots as shown in Fig.~\ref{fig:quantumdotmodel}. $\uparrow$ and $\downarrow$ represents spins, and  $\mathcal{I}_{N,\mathrm{S}}$ and $\mathcal{I}_{N,\mathrm{T}}$ are factors related to normalization, given in Appendix \ref{Expression}.

In the biased DQD, the singlet states hybridize as
\begin{equation}
	\label{eq:singlet02n}
	|\mathrm{S}_{\rm mix}^{(0,2N)}\rangle=\frac{|\mathrm{S}(1,2N-1)\rangle+ \beta|\mathrm{S}(0,2N)\rangle}{\sqrt{1+\beta^{2}}},
\end{equation}
and
\begin{equation}
	\label{eq:singlet22n2}
	|\mathrm{S}_{\rm mix}^{(2,2N-2)}\rangle=\frac{|\mathrm{S}(1,2N-1)\rangle+ \beta|\mathrm{S}(2,2N-2)\rangle}{\sqrt{1+\beta^{2}}},
\end{equation}
where
\begin{equation}
	|\mathrm{S}(0,2N)\rangle=|\uparrow_{\mathrm{R}_{N}}\downarrow_{\mathrm{R}_{N}}\uparrow_{\mathrm{R}_{N-1}}\downarrow_{\mathrm{R}_{N-1}}...\uparrow_{\mathrm{R}_{1}}\downarrow_{\mathrm{R}_{1}}\rangle,
\end{equation}
\begin{equation}
	|\mathrm{S}(2,2N-2)\rangle=|\uparrow_{\textrm{L}_{1}}\downarrow_{\textrm{L}_{1}}\uparrow_{\mathrm{R}_{N-1}}\downarrow_{\mathrm{R}_{N-1}}...\uparrow_{\textrm{R}_{1}}\downarrow_{\textrm{R}_{1}}\rangle.
\end{equation}
Here, $1/\sqrt{1+\beta^2}$ and $\beta/\sqrt{1+\beta^2}$ are both functions of detuning that can be calculated through the Hund-Mulliken approximation.

\subsection{Multi-electron dephasing of electron-phonon interaction}
In a semiconductor, the Hamiltonian that describes effective electron-phonon interaction takes the form:
\begin{equation}
	H_\mathrm{ep}=\sum_{\textbf{q},\lambda }^{}M_{\lambda  }(\textbf{q})\rho (\textbf{q})\sigma _{z}(a_{\textbf{q},\lambda}+a_{\textbf{q},\lambda}^{\dagger }),
\end{equation}
where $a_{\textbf{q},\lambda}$ and $a_{\textbf{q},\lambda}^{\dagger }$ are phonon annihilation and creation operators respectively, $\textbf{q}$ the lattice momentum, and $\lambda$ the branch index. $M(\textbf{q})$ represents different kinds of electron-phonon interactions. 
In GaAs DQD, the deformation potential (DP) and piezoelectric (PE) interaction provides the main contribution to the phonon dephasing, while contributions from other interactions are negligible  \cite{PhysRevB.83.165322,chan2022microscopic}. The DP and PE have the form 
\begin{equation}
	M_\mathrm{GaAs}^\mathrm{DP}(\textbf{q})=D\left(\frac{\hbar}{\rho V\omega _{\textbf{q}}}\right)^{\frac{1}{2}}|\textbf{q}|,
\end{equation}
\begin{equation}
    M_\mathrm{GaAs}^\mathrm{PE}(\textbf{q})=i(\frac{\hbar}{\rho V\omega _{\textbf{q}}})^{\frac{1}{2}}2ee_{14}(\hat{q}_{x}\hat{q}_{y}\hat{\xi }_{z}+\hat{q}_{y}\hat{q}_{z}\hat{\xi }_{x}+\hat{q}_{z}\hat{q}_{x}\hat{\xi }_{y}),
\end{equation}
and one should note that $M_\mathrm{GaAs}^\mathrm{DP}(\textbf{q})$  only couples  electrons to longitudinal acoustic phonons and $ M_\mathrm{GaAs}^\mathrm{PE}(\textbf{q})$ can couple electrons to both LA and transverse acousitc phonons. Here, $D = 8.6$ eV is the deformation constant, $\rho$ = $5.3\times  10^{3}$ kg/m$^3$ the mass density, $e$ is elementary electric charge, $e_{14}=1.38\times  10^{9}$ $V/m$ is elasticity tensor component, $\hat{\xi }$ is the polarization vector, and $\omega _{\textbf{q}}$ the angular frequency of the phonon mode $\textbf{q}$. We further define $\gamma_{\textbf{q}}$ as the population relaxation rate of the phonon mode $\textbf{q}$, which is assumed to have the form $\gamma_{\textbf{q}}=\gamma_{0}q^{n}$ in our calculations. We fix $\gamma_{0}=10^8$ Hz and consider cases in which $n=2$ or $n=3$ \cite{PhysRevB.83.165322}, and we have also verified that other values of $\gamma_0$ and $n$ will not  significantly change our main findings.

The off-diagonal element  of the effective electron-phonon interaction Hamiltonian leads to a decay in the form
\begin{equation}
	\rho _\mathrm{ST}(t)=\rho _\mathrm{ST}(0)e^{-B^{2}(t)},
\end{equation}
where $B^{2}(t)$ is dephasing factor. For a dissipative phonon reservoir with finite $\gamma_{\textbf{q}}$, the main contribution to $B^{2}(t)$ can be calculated by \cite{PhysRevB.83.165322}
\begin{equation}
	\label{eq:B2t}
	B_{\rm Decay}^{2}(t)=\frac{V}{2\pi ^{3}\hbar^{2}}\int d^{3}\textbf{q}\frac{|M(\textbf{q})A_{\phi}(\textbf{q})|^{2}}{\omega_{\textbf{q}}^{2}+\gamma_{\textbf{q}}^{2}/4}\frac{\gamma_{\textbf{q}}}{2}t\equiv\Gamma _{\textrm{ST}}t,
\end{equation}
where $\Gamma _{\textrm{ST}}$, the dephasing rate, is the key quantity considered in this paper, 
and 
$A_{\phi }$ is given by
\begin{equation}
	\label{eq:Aphi}
	A_{\phi }=\frac{1}{2}[\langle \psi_{\textrm{T}}|\rho (\mathbf{q})|\psi_{\textrm{T}} \rangle-\langle \psi_{\textrm{S}}|\rho (\mathbf{q})|\psi_{\textrm{S}} \rangle]\equiv A_{\phi }(\mathbf{q_{||}})f(q_{z}),
\end{equation}
where $\psi_{\textrm{T}}$ is the triplet state of Eq.~\eqref{eq:triplet2n1}, and $\psi_{\textrm{S}}$
is the singlet state from Eq.~\eqref{eq:singlet2n1}
in unbiased case, and is the state from Eq.~\eqref{eq:singlet02n} or Eq.~\eqref{eq:singlet22n2}
 in biased case.
   $\rho (\mathbf{q})$ is the electron density operator, taking the form $\rho (\mathbf{q})=\sum_{i=1}^{2N}e^{i\, \mathbf{q\cdot \mathrm{R}_{i}}}$. $A_{\phi }(\mathbf{q_{||}})$ is obtained from the $x$ and $y$ components of orbital states, and $f(q_{z})$ is solely determined by the $z$-direction wave function, given by
\begin{equation}
	f(q_{z})=\frac{\sin\left(q_{z}a_{z}\right)}{q_{z}a_{z}}\frac{-\pi ^{2}}{(q_{z}a_{z})^{2}-\pi ^{2}},
\end{equation}
where $q_{z}$ is $z$-component lattice momentum, $a_{z}$ = $3\times 10^{-9}\mathrm{m}$ is width of the infinite square well for acoustic phonons. Details on the evaluation of  $A_{\phi }$  is given in the Appendix  \ref{Expression}.

\section{Results}
\label{sec:result}
\subsection{Dephasing rate of unbiased case}

\begin{figure}
	\includegraphics[scale=0.45]{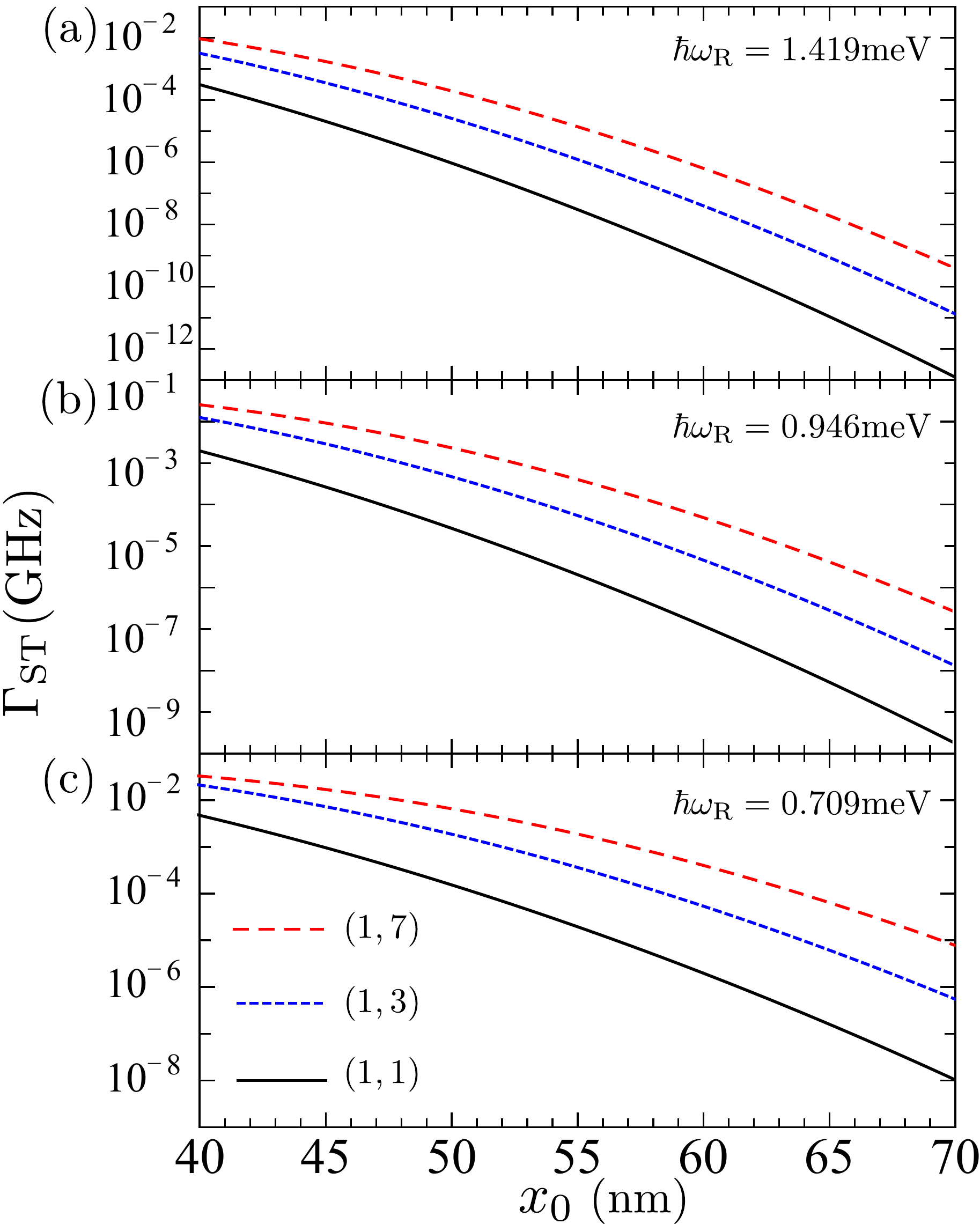}
	\caption{The dephasing rate $\Gamma _{\textrm{ST}}$ v.s.~half dot distance $x_{0}$ in unbiased case for the three different electron configurations as indicated and the right dot confinement energy $\hbar \omega _{\textrm{R}}$ being (a) 1.419 meV, (b) 0.946 meV and (c) 0.709 meV. The left dot confinement energy is fixed as $\hbar \omega _{\textrm{L}}=2.838 $ meV.  }\label{fig:Fig2}
\end{figure}

According to Eq.~\eqref{eq:B2t}, the electron-phonon dephasing rate can be expressed as
\begin{equation}
	\Gamma_\mathrm{ST}=\frac{V}{2\pi ^{3}\hbar^{2}}\int d^{3}\textbf{q}\frac{|M(\textbf{q})A_{\phi}(\textbf{q})|^{2}}{\omega_{\textbf{q}}^{2}+\gamma_{\textbf{q}}^{2}/4}\frac{\gamma_{\textbf{q}}}{2}.
\end{equation}
In the unbiased case, $A_\phi$ is dependent on the singlet state Eq.~\eqref{eq:singlet2n1} and triplet state Eq.~\eqref{eq:triplet2n1}, suggesting that $\Gamma_\mathrm{ST}$ varies with the number of electrons.  

Here we consider three unbiased cases with electron configurations 
 $(1,1)$, $(1,3)$, and  $(1,7)$ with the first entry showing the number of electron in dot $\mathrm{L}$ and the second dot $\mathrm{R}$. A schematic showing the latter two cases is shown in Fig.~\ref{fig:quantumdotmodel}(b).
Details on the evaluation of $A_{\phi}$ in these cases are given in Appendix \ref{Expression}.

Figure~\ref{fig:Fig2} shows the dephasing rate $\Gamma_\mathrm{ST}$ as functions of the half dot distance $x_{0}$ with different confinement strength $\hbar \omega _{\textrm{R}}$ as indicated. The three values of the confinement strength on dot R $\hbar \omega _{\textrm{R}}=1.419$ meV, 0.946 meV, and 0.709 meV correspond to dot sizes $28.076$ nm, $33.981$ nm, and $38.627$ nm, respectively. Several features can be clearly seen from the figure. Firstly, the dephasing rate rapidly decreases with increasing $x_0$ in all cases. The results for $(1,1)$ are consistent with Ref.~\cite{PhysRevB.83.165322}, and it is not surprising that results for $(1,3)$ and $(1,7)$ are similar. Secondly, for a given confinement strength, the dephasing rate is greatest for $(1,7)$ as more electrons imply larger integration from $A_{\phi}$ as Appendix \ref{Expression} shows, implying more channels of electron-phonon interaction. Similarly, the effect is 
 intermediate for $(1,3)$, and smallest for $(1,1)$. Thirdly, when $x_0$ and $\hbar\omega_\mathrm{L}$ are fixed, the dephasing rate is greater when dot R is larger (smaller $\hbar\omega_\mathrm{R}$) and smaller when dot R is smaller  (larger $\hbar\omega_\mathrm{R}$). As can be seen from Eq.~\eqref{eq:aphiii}, the behavior of $A_{\phi}$ is controlled by integrals Eq.~\eqref{eq:i1}, Eq.~\eqref{eq:i2} and Eq.~\eqref{eq:i4}, for the cases of  (1,1), (1,3), and (1,7), respectively. The l.h.s. of Eq.~\eqref{eq:i1}, Eq.~\eqref{eq:i2} and Eq.~\eqref{eq:i4} decreases either as $x_0$ increase, or as dots get smaller.

\subsection{Biased case and the merit figure}

\begin{figure}
	\includegraphics[scale=0.60]{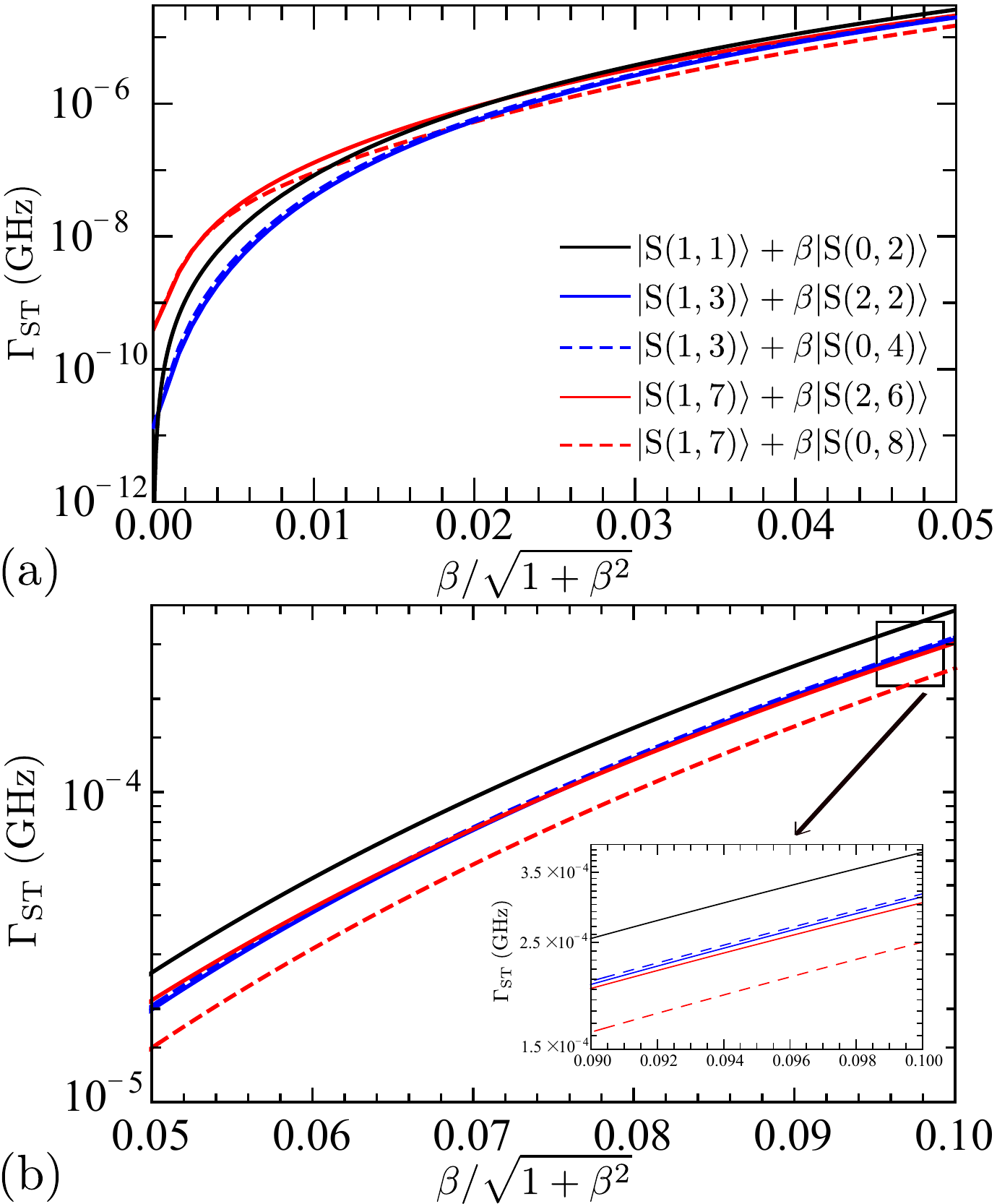}
	\caption{The dephasing rate $\Gamma _{\textrm{ST}}$ v.s.~$\beta/\sqrt{1+\beta^2}$ in biased case for five different states as indicated (note that the normalization constant is omitted in the legend). (a) shows the range 
$0\le\beta/\sqrt{1+\beta^2}\le0.05$, and (b) the range $0.05\le\beta/\sqrt{1+\beta^2}\le0.1$, with an inset showing the zoomed-in version at the tail of the curves. Parameters: $x_{0}=70  $ nm, $\hbar \omega _{\textrm{L}}=2.838 $ meV, and $\hbar \omega _{\textrm{R}}=1.419 $ meV.}
	\label{fig:Fig3}
\end{figure}

\begin{figure}
	\includegraphics[scale=0.475]{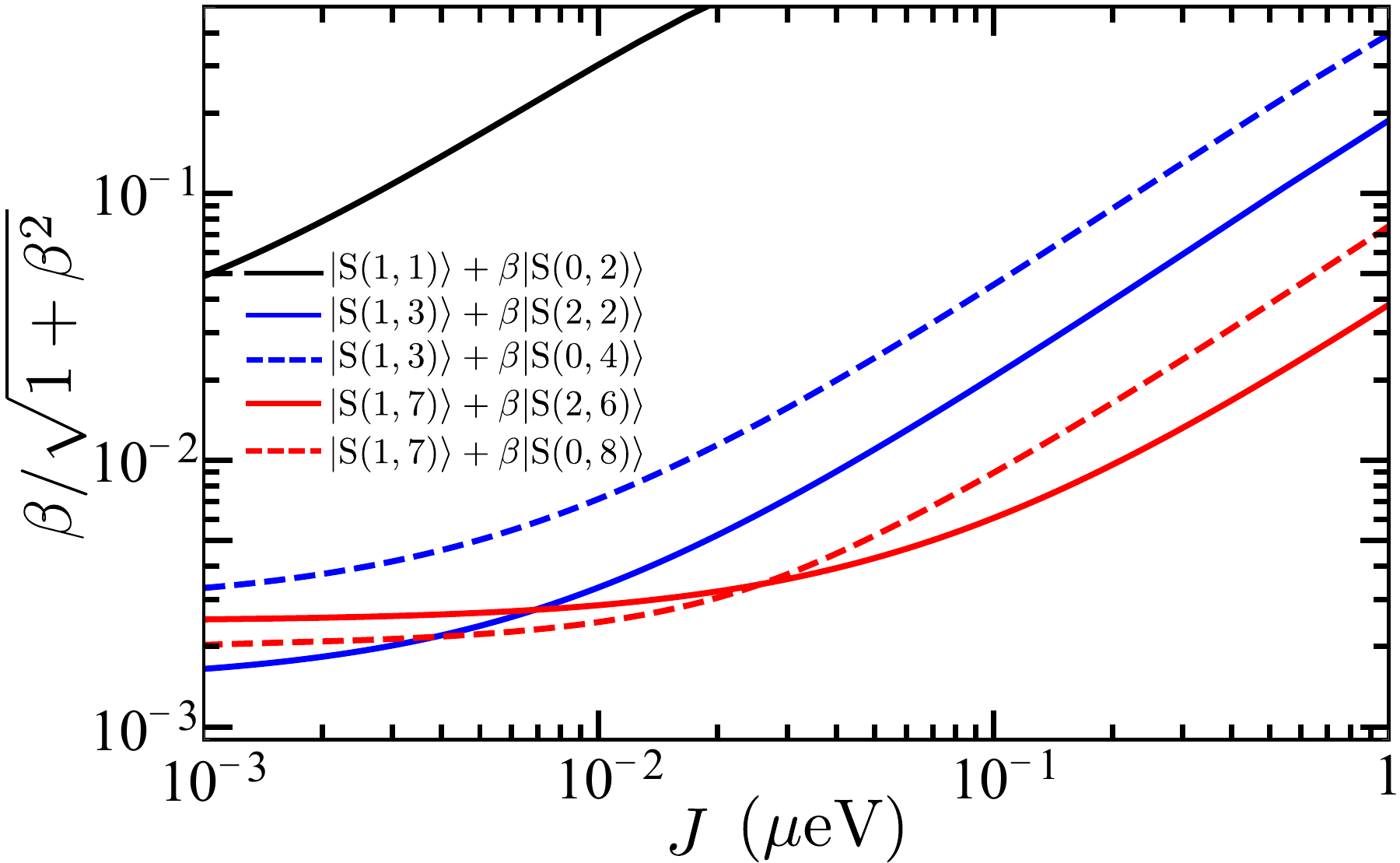}
	\caption{$\beta/\sqrt{1+\beta^2}$ as functions of the exchange energy $\textit{J}$ in the biased case for five different states as indicated. Parameters: $x_{0}=70$ nm,  $\hbar \omega _{\textrm{L}}=2.838 $ meV,  $\hbar \omega _{\textrm{R}}=1.419 $ meV. }
	\label{fig:Fig4} 
\end{figure}

\begin{figure}
\includegraphics[scale=0.50]{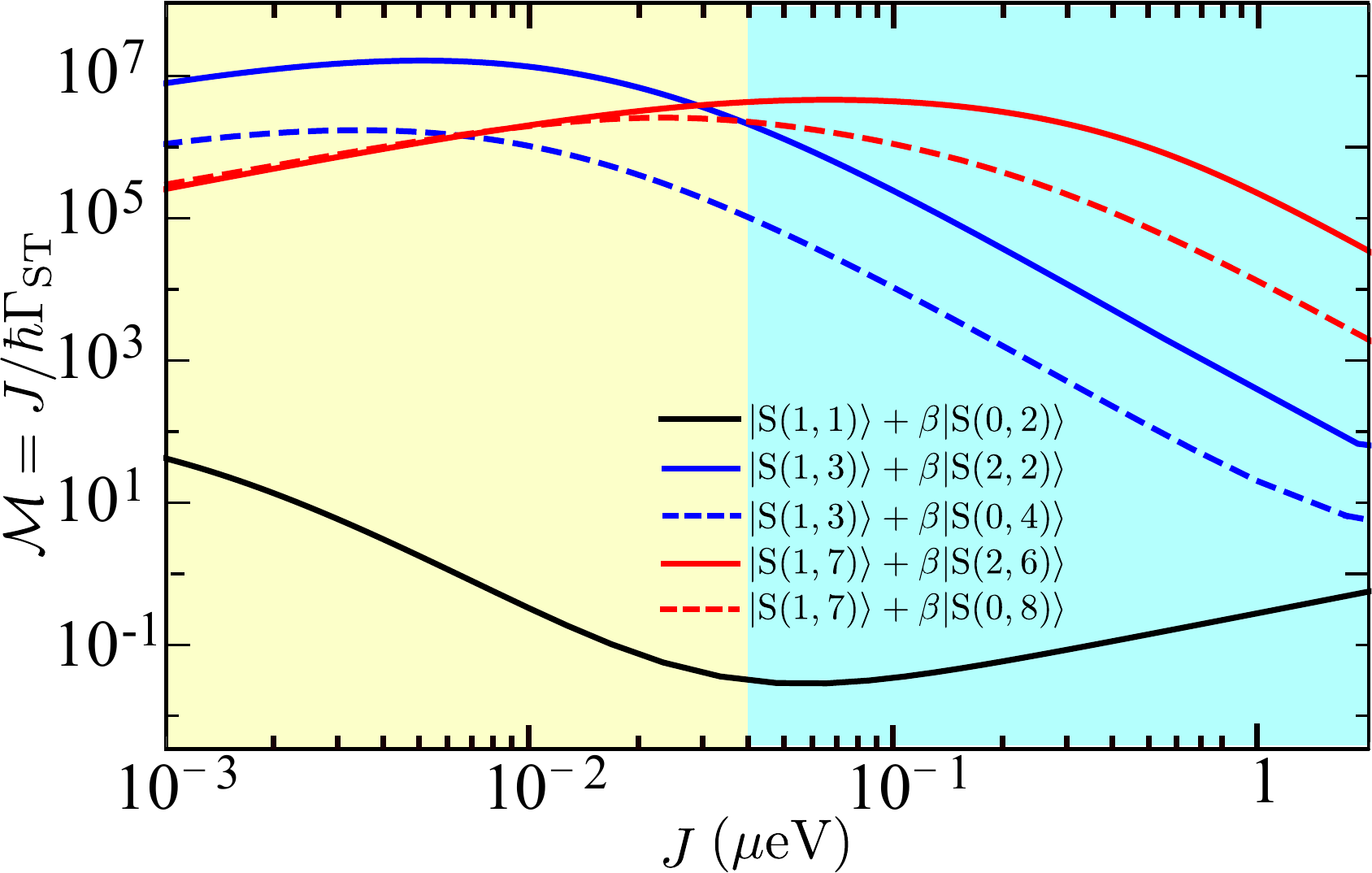}
\caption{The merit figure v.s.~exchange energy for five different states as indicated. Parameters: half dot distance $x_{0}=70 $ nm,  $\hbar \omega _{\textrm{L}}=2.838 $ meV,  and $\hbar \omega _{\textrm{R}}=1.419 $ meV. The yellow shaded area ($J<0.04\ \mu$eV) shows the regime where the merit figure of (1,3) and (1,7) are better than (1,1), while the cyan shaded area ($J\ge0.04\ \mu$eV ) the regime where the merit figure for (1,7) is the greatest, that for (1,3) is intermediate and for (1,1) is the lowest.}
\label{fig:Fig5}\end{figure}

As the detuning $\Delta$ changes, the singlet states start to hybridize as suggested by Eq.~\eqref{eq:singlet02n} or Eq.~\eqref{eq:singlet22n2}. Fig.~\ref{fig:Fig3} shows the dephasing rate $\Gamma_\mathrm{ST}$ as functions of $\beta/\sqrt{1+\beta^2}$ for six different hybridized states as indicated (the normalization constant is ignored in the legend). Note that for small $\beta$, what we call the hybridization ratio $\beta/\sqrt{1+\beta^2}\approx\beta$ indicates the ratio of the hybridization to states other than the $(1,1)$, $(1,3)$, and $(1,7)$ states considered. 
Fig.~\ref{fig:Fig3}(a) shows the range $0<\beta/\sqrt{1+\beta^2}<0.05$ while Fig.~\ref{fig:Fig3}(b) the range $0.05<\beta/\sqrt{1+\beta^2}<0.10$. We can see that while $\Gamma_\mathrm{ST}$ increases monotonically with the hybridization ratio, the order of the results for states with mainly $(1,1)$, $(1,3)$, and $(1,7)$ character changes. In particular, for $\beta/\sqrt{1+\beta^2}=0$ the dephasing rate for the state $(1,1)$ is the smallest, consistent with the unbiased case. However, for $\beta/\sqrt{1+\beta^2}\gtrsim0.02$ the dephasing rate for the state with mainly $(1,1)$ character becomes the largest, which is greater than the case with mainly $(1,3)$ character by  about 30\% and the case with mainly $(1,7)$ character by about 60\% at  $\beta/\sqrt{1+\beta^2}\approx0.1$, as can be seen from the inset of Fig.~\ref{fig:Fig3}(b). This results is opposite to the unbiased case and is a direct consequence of the state hybridization.

Figure~\ref{fig:Fig4} shows the hybridization ratio $\beta/\sqrt{1+\beta^2}$ versus the exchange interaction $J$ as calculated from Eq.~\eqref{eq:defJ} using the Hund-Mulliken method. In general, the more hybridized the singlet state is, the larger the absolute value of detuning should be, and as a consequence, the greater the value of $J$ is.  For the same value of $J$, the state with mainly $(1,1)$ character has a greater hybridization ratio \cite{PhysRevB.91.035301}.

To reveal the performance of our system in realistic situations, we define the merit $\mathcal{M}=J/\hbar\Gamma _{\textrm{ST}}$ as the ratio between the exchange gate time given by $\hbar/J$ and the decay time given by $1/\Gamma _{\textrm{ST}}$. The results of the merit $\mathcal{M}$ as functions of the exchange energy is shown in Fig.~\ref{fig:Fig5}, which is the key result of this paper. The non-monotonic behavior of the $\mathcal{M}$ v.s.~$J$ curves shown in Fig.~\ref{fig:Fig5} is a combinational effect of the changing rate of $\Gamma _{\textrm{ST}}$ and $J$ as functions of $\beta/\sqrt{1+\beta^2}$.  More importantly, the merit figure for states associated with $(1,3)$ and $(1,7)$ are greater than those associated with $(1,1)$. This indicates that multi-electron quantum dots may offer advantages in electron-phonon dephasing, which is the main result of this paper. The results shown in Fig.~\ref{fig:Fig5} is divided into two regions: $J<0.04\ \mu$eV (marked by yellow), and  $J\ge0.04\ \mu$eV (marked by cyan). In the right (cyan) region, the merit figures for states associated with $(1,7)$ are greater than those for $(1,3)$, while the merit figure for the state with $(1,1)$ is the smallest. Given the fact that in practical operations of the qubit, the exchange interaction should neither be too small nor too large. Therefore, in the regime of $J\ge 0.04\ \mu\mathrm{eV}$, having more electrons in the right dot implies a better merit figure, which is advantageous in experiments. This is the key finding of this paper. We have also verified that our conclusion holds for other experimentally relevant parameters, including the dot distance and dot sizes (confinement strength), and selective results are shown in
Appendix \ref{Merit figures of other quantum dot parameters}. This behavior is understandable from Fig.~\ref{fig:Fig3} and Fig.~\ref{fig:Fig4}. From Fig.~\ref{fig:Fig4}, one sees that at a fixed value of $\beta/\sqrt{1+\beta^2}$, $J$ is the largest for states associated with $(1,7)$, intermediate for $(1,3)$ and smallest for $(1,1)$, and the differences between them are quite appreciable. On the other hand, from Fig.~\ref{fig:Fig3} one sees that for the same value of $\beta/\sqrt{1+\beta^2}$, the values of $\Gamma_\mathrm{ST}$ are close. Since $J$ is on the numerator of the merit figure, the merit figure should follow the same trend as observed in Fig.~\ref{fig:Fig4}.

\section{Conclusions}
\label{sec:conclusion}

In this paper, we have calculated the dephasing rate, exchange energy and the merit figure of a multi-electron quantum-dot system with one electron in the  left dot and 1,3 or 7 electrons in the right  dot. We have found that in the unbiased case, the dephasing rate in general increases with the number of electrons in the right dot. This is however not necessarily true in the biased case. Nevertheless, as we have shown that in the experimentally relevant regime $J\ge 0.04\ \mu\mathrm{eV}$, having more electrons in the right dot implies a better merit figure. Our results suggest that  multi-electron quantum dots may be advantageous in certain cases.


\section*{Acknowledgement}

This work is supported by the Key-Area Research and Development Program of GuangDong Province  (Grant No. 2018B030326001), the National Natural Science Foundation of China (Grant No. 11874312), the Research Grants Council of Hong Kong (Grant No. CityU 11303617), and the Guangdong Innovative and Entrepreneurial Research Team Program (Grant No. 2016ZT06D348).


\appendix

\section{Expression of $A_{\phi }$}
\label{Expression}

The singlet and triplet states of unbiased case can be written as
\begin{equation}
	|\mathrm{S}\rangle=a(|\uparrow_{\mathrm{L}_{1}}\downarrow_{\mathrm{R}_{N}}...\uparrow_{\mathrm{R}_{1}}\downarrow_{\mathrm{R}_{1}}\rangle+|\uparrow_{\mathrm{R}_{N}}\downarrow_{\mathrm{L}_{1}}...\uparrow_{\mathrm{R}_{1}}\downarrow_{\mathrm{R}_{1}}\rangle),
\end{equation}
and
\begin{equation}
	|\mathrm{T}\rangle=b(|\uparrow_{\mathrm{L}_{1}}\downarrow_{\mathrm{R}_{N}}...\uparrow_{\mathrm{R}_{1}}\downarrow_{\mathrm{R}_{1}}\rangle-|\uparrow_{\mathrm{R}_{N}}\downarrow_{\mathrm{L}_{1}}...\uparrow_{\mathrm{R}_{1}}\downarrow_{\mathrm{R}_{1}}\rangle),
\end{equation}
where $|\mathrm{S}\rangle$ and $|\mathrm{T}\rangle$ satisfy 
\begin{equation}
	\langle\mathrm{S}|\mathrm{S}\rangle=1,
	\langle\mathrm{T}|\mathrm{T}\rangle=1,
	\langle\mathrm{S}|\mathrm{T}\rangle=0,
\end{equation}
therefore we can obtain
\begin{equation}
	a=\frac{1}{\sqrt{2(1+\mathcal{I}_{N,\mathrm{S}})}},
	b=\frac{1}{\sqrt{2(1-\mathcal{I}_{N,\mathrm{T}})}},
\end{equation}
as indicated in Eq.~\eqref{eq:singlet2n1} and Eq.~\eqref{eq:triplet2n1}. $\mathcal{I}_{N,\mathrm{S}}$ and $\mathcal{I}_{N,\mathrm{T}}$ are factors dependent on electron numbers to be calculated below.

We denote $I_{i}=\langle\mathrm{\mathrm{L}_{1}}|\mathrm{\mathrm{R}_{i}}\rangle=\langle\mathrm{\mathrm{R}_{i}}|\mathrm{\mathrm{L}_{1}}\rangle$, where $1\leq i,j<N$. $\mathrm{L}_{1}$ and $\mathrm{R}_{N}$ are wave functions (without spin part) based on Fock-Darwin states.

For $N>1$, by applying the Slater-Condon rules \cite{szabo1982modern,PhysRevB.84.235309}, we have
\begin{equation}
	\mathcal{I}_{N,\mathrm{S}}=I_{N}^{2}-\sum_{i=1}^{N-1}I_{i}^{2},
\end{equation}
and
\begin{equation}
	\mathcal{I}_{N,\mathrm{T}}=I_{N}^{2}+\sum_{i=1}^{N-1}I_{i}^{2}.
\end{equation}

For unbiased case, we can express $A_{\phi}$ as
\begin{eqnarray}
	A_{\phi }&=&\frac{1}{2}[\langle \mathrm{T}(1,2N-1)|\rho (\mathbf{q})|\mathrm{T}(1,2N-1) \rangle\notag\\
	&&-\langle \mathrm{S}(1,2N-1)|\rho (\mathbf{q})|\mathrm{S}(1,2N-1) \rangle]
	\notag\\
	&=&A_{\phi }(\mathbf{q_{||}})f(q_{z}).
\end{eqnarray} 
For $N=1$, the expression of $A_{\phi}$ has been explicitly shown in \cite{PhysRevB.83.165322}. Here, we give a general expression of $A_\phi$ for $N>1$: 
\begin{equation}
	\langle \mathrm{S}(1,2N-1)|\rho (\mathbf{q})|\mathrm{S}(1,2N-1) \rangle
	=\frac{\varrho_{+}}{1+\mathcal{I}_{N,\mathrm{S}}},
\end{equation}
\begin{equation}
	\langle \mathrm{T}(1,2N-1)|\rho (\mathbf{q})|\mathrm{T}(1,2N-1) \rangle
	=\frac{\varrho_{-}}{1-\mathcal{I}_{N,\mathrm{T}}},
\end{equation}
where
\begin{widetext}
	\begin{align}
		\varrho_{\pm }&=\rho_{\textrm{L}_{1},\textrm{L}_{1}}+\rho_{\textrm{R}_{\textrm{N}},\textrm{R}_{\textrm{N}}}+2\sum_{\textrm{i}=1}^{{N-1}}\rho_{\textrm{R}_{\textrm{i}},\textrm{R}_{\textrm{i}}}\pm I_{{N}}(\rho_{\textrm{L}_{1},\textrm{R}_{{N}}}+\rho_{\textrm{R}_{{N}},\textrm{L}_{1}})-\sum_{\textrm{i}=1}^{{N}-1}\left [I_{i}(\rho_{\textrm{L}_{1},\textrm{R}_{i}}+\rho_{\textrm{R}_{\textrm{i}},\textrm{L}_{1}})+I_{\textrm{i}}^{2}(\rho_{\textrm{R}_{\textrm{i}},\textrm{R}_{\textrm{i}}}+\rho_{\textrm{R}_{{N}},\textrm{R}_{{N}}})\right]
		\notag\\
		&\quad\mp \sum_{\textrm{i}=1}^{{N-1}}(-2I_{{N}}^{2}\rho_{\textrm{R}_{\textrm{i}},\textrm{R}_{\textrm{i}}}+I_{\textrm{i}}I_{{N}}\rho_{\textrm{R}_{\textrm{i}},\textrm{R}_{{N}}}+I_{{N}}I_{\textrm{i}}\rho_{\textrm{R}_{{N}},\textrm{R}_{\textrm{i}}})+\sum_{\textrm{j}=1}^{{N}-1}\sum_{\textrm{i}=1,\textrm{i}\neq \textrm{j}}^{{N}-1}(I_{\textrm{i}}I_{\textrm{j}}\rho_{\textrm{R}_{\textrm{i}},\textrm{R}_{\textrm{j}}}-2I_{\textrm{j}}^{2}\rho_{\textrm{R}_{\textrm{i}},\textrm{R}_{\textrm{i}}}).
	\end{align}
\end{widetext}
Here, $I_{i}=\langle\mathrm{\mathrm{L}_{1}}|\mathrm{\mathrm{R}_{i}}\rangle$, $\rho_{\mathrm{R}_{i},\mathrm{R}_{j}}=\langle\mathrm{\mathrm{R}_{i}}|\rho |\mathrm{\mathrm{R}_{j}}\rangle$, and similarily,  $\rho_{\mathrm{L}_{1},\mathrm{R}_{j}}=\langle\mathrm{\mathrm{L}_{1}}|\rho |\mathrm{\mathrm{R}_{i}}\rangle$. We then have
\begin{equation}
	\label{eq:aphiii}
	A_{\phi }=\frac{2I_{{N}}\mathbb{I}_{1}(1-\sum_{i=1}^{N-1}I_{i}^{2})+2I_{{N}}^{2}\mathbb{I}_{2}}{1-I_{N}^{4}-(2-\sum_{i=1}^{N-1}I_{i}^{2})\sum_{i=1}^{N-1}I_{i}^{2}},  
\end{equation}
\begin{equation}
	\mathbb{I}_{1}=-(\rho_{\textrm{L}_{\textrm{1}},\textrm{R}_{N}}+\rho_{\textrm{R}_{\textrm{N}},\textrm{L}_{1}})+\sum_{\textrm{i}=1}^{{N}-1}I_{i}(\rho_{\textrm{R}_{N},\textrm{R}_{i}}+\rho_{\textrm{R}_{\textrm{i}},\textrm{R}_{N}}),
	\label{eq:II1}
\end{equation}

and
\begin{equation}
	\begin{split}
		\mathbb{I}_{2}=&\rho_{\textrm{L}_{1},\textrm{L}_{1}}+\rho_{\textrm{R}_{N},\textrm{R}_{N}}+2\sum_{i=1}^{N-1}I_{i}^{2}\sum_{\textrm{i}=1}^{{N-1}}\rho_{\textrm{R}_{\textrm{i}},\textrm{R}_{\textrm{i}}}\\
		&-\sum_{\textrm{i}=1}^{N-1}\left[I_{i}(\rho_{\textrm{L}_{1},\textrm{R}_{i}}+\rho_{\textrm{R}_{\textrm{i}},\textrm{L}_{1}})+I_{\textrm{i}}^{2}(\rho_{\textrm{R}_{\textrm{i}},\textrm{R}_{\textrm{i}}}+\rho_{\textrm{R}_{{N}},\textrm{R}_{{N}}})\right]\\
		&+\sum_{\textrm{j}=1}^{{N}-1}\sum_{\textrm{i}=1,\textrm{i}\neq \textrm{j}}^{{N}-1}(I_{\textrm{i}}I_{\textrm{j}}\rho_{\textrm{R}_{\textrm{i}},\textrm{R}_{\textrm{j}}}-2I_{\textrm{j}}^{2}\rho_{\textrm{R}_{\textrm{i}},\textrm{R}_{\textrm{i}}}).
	\end{split}\label{eq:II2}
\end{equation}
Here, it is straightforward to show that, for $i<j$, we have $I_{i}\ll 1$, $I_{i}\ll I_{j},  \rho_{\textrm{R}_{i}, \textrm{R}_{i}}<\rho_{\textrm{R}_{j}, \textrm{R}_{j}}, \rho_{\textrm{L}_{1},\textrm{R}_{i}}<\rho_{\textrm{L}_{1},\textrm{R}_{j}}\ $. For (1,1),(1,3) and (1,7), we have $N=1,2,4$, therefore
\begin{equation}
	\label{eq:i1}
	I_{1}=2e^{-2x_{0}^{2}/(l_{L}^{2}+l_{R}^{2})}l_{L}l_{R}/(l_{L}^{2}+l_{R}^{2}),
\end{equation}
\begin{equation}
	\label{eq:i2}
	I_{2}=4x_{0}e^{-2x_{0}^{2}/(l_{L}^{2}+l_{R}^{2})}l_{L}l_{R}^{2}/(l_{L}^{2}+l_{R}^{2})^{2},
\end{equation}
\begin{equation}
	\label{eq:i4}
	I_{4}=4\sqrt{2}x_{0}^{2}e^{-2x_{0}^{2}/(l_{L}^{2}+l_{R}^{2})}l_{L}l_{R}^{3}/(l_{L}^{2}+l_{R}^{2})^{3},
\end{equation}
where $l_{L}$ is left dot cofinement length and $l_{R}$ is right dot cofinement length that can be calculated from their confinement strength.  Therefore in Eq.~\eqref{eq:aphiii} numerator,  $I_{\textrm{N}}^{2}\mathbb{I}_{2}\ll I_{\textrm{N}}\mathbb{I}_{1}$. As $N$ increases, $\mathbb{I}_{1}$ and $I_{\textrm{N}}$ also increases, eventually lead to increases of  $\left |A_{\phi }  \right |^{2}$ and the dephasing rate. One can also find that due to $A_{\phi }\sim I_{\textrm{N}}$, therefore $\left |A_{\phi }  \right |^{2}$ decreases as $x_{0}$  and $\hbar \omega _{\textrm{R}}$ increase.

In biased case, the explicit expression of $A_{\phi }$ at $N=1$ can also be found in \cite{PhysRevB.83.165322}. For $N>1$, there are two situations of biased case in our consideration.  From Eq.~\eqref{eq:singlet02n}, we have
\begin{widetext}
	\begin{eqnarray}
		A_{\phi }&=&\frac{1}{2}[\langle \mathrm{T}(1,2N-1)|\rho (\mathbf{q})|\mathrm{T}(1,2N-1) \rangle-\langle\mathrm{S}_{\rm mix}^{(0,2N)}|\rho (\mathbf{q})|\mathrm{S}_{\rm mix}^{(0,2N)}\rangle]
		\notag\\
		&=&\frac{1}{2}[\langle \mathrm{T}(1,2N-1)|\rho (\mathbf{q})|\mathrm{T}(1,2N-1) \rangle-[\langle \mathrm{S}(1,2N-1)|\rho (\mathbf{q})|\mathrm{S}(1,2N-1)\rangle-2\beta\langle \mathrm{S}(1,2N-1)|\rho (\mathbf{q})|\textbf{S}(0,\textrm{2N})\rangle\notag\\
		&&-\beta ^{2}\langle \mathrm{S}(0,\textrm{2N})|\rho (\mathbf{q})|\mathrm{S}(0,\textrm{2N})\rangle]/(1+\beta^{2})],\notag\\
	\end{eqnarray}
	and from Eq.~\eqref{eq:singlet22n2}, we have
	\begin{eqnarray}
		A_{\phi }&=&\frac{1}{2}[\langle \mathrm{T}(1,2N-1)|\rho (\mathbf{q})|\mathrm{T}(1,2N-1) \rangle-\langle\mathrm{S}_{\rm mix}^{(0,2N)}|\rho (\mathbf{q})|\mathrm{S}_{\rm mix}^{(0,2N)}\rangle]\notag\\
		&=&\frac{1}{2}[\langle \mathrm{T}(1,2N-1)|\rho (\mathbf{q})|\mathrm{T}(1,2N-1) \rangle-[\langle \mathrm{S}(1,2N-1)|\rho (\mathbf{q})|\mathrm{S}(1,2N-1)\rangle-2\beta\langle \mathrm{S}(1,2N-1)|\rho (\mathbf{q})|\textbf{S}(2,\textrm{2N-2})\rangle\notag\\
		&&-\beta ^{2}\langle \mathrm{S}(2,\textrm{2N-2})|\rho (\mathbf{q})|\mathrm{S}(2,\textrm{2N-2})\rangle]/(1+\beta^{2})],\notag\\
	\end{eqnarray}
	where
	\begin{equation}
		\langle \mathrm{S}(0,\textrm{2N})|\rho (\mathbf{q})|\mathrm{S}(0,\textrm{2N})\rangle]=2\sum_{i=1}^{{N}}\rho_{\textrm{R}_{\textrm{i}},\textrm{R}_{\textrm{i}}},
	\end{equation}
	
	\begin{eqnarray}
		\langle \mathrm{S}(0,\textrm{2N})|\rho (\mathbf{q})|\mathrm{S}1,2N-1)\rangle]=\frac{1}{\sqrt{2(1+\mathcal{I}_{N,\mathrm{S}})}}\left(4I_{{N}}\sum_{\textrm{i}=1}^{{N}-1}\rho_{\textrm{R}_{\textrm{i}},\textrm{R}_{\textrm{i}}}-2\sum_{\textrm{i}=1}^{{N}-1}I_{\textrm{i}}\rho_{\textrm{R}_{\textrm{i}},\textrm{R}_{\textrm{i}}}+2I_{{N}}\rho_{\textrm{R}_{{N}},\textrm{R}_{{N}}}\right),\notag\\
	\end{eqnarray}
	
	\begin{equation}
		\begin{split}
			\langle \mathrm{S}(2,\textrm{2N-2})|\rho (\mathbf{q})|\mathrm{S}2,\textrm{2N-2})\rangle]=&
			2\rho_{\textrm{L}_{1},\textrm{L}_{1}}\left(1-\sum_{\textrm{i}=1}^{{N}-1}I_{\textrm{i}}^{2}\right) + 2\sum_{\textrm{i}=1}^{{N-1}}\left [ \rho_{\textrm{R}_{\textrm{i}},\textrm{R}_{\textrm{i}}}\left(1-\sum_{\textrm{i}=1}^{{N}-1}I_{\textrm{i}}^{2}\right)-I_{\textrm{i}}(\rho_{\textrm{L}_{1},\textrm{R}_{\textrm{i}}}+\rho_{\textrm{R}_{\textrm{i}},\textrm{L}_{1}}) \right ]\\
			&+\sum_{\textrm{i},\textrm{j}\neq \textrm{i}}^{{N}-1}\left(2I_{\textrm{i}}I_{\textrm{j}}\rho_{\textrm{R}_{\textrm{i}},\textrm{R}_{\textrm{j}}}-4I_{\textrm{i}}^{2}\rho_{\textrm{R}_{\textrm{i}},\textrm{R}_{\textrm{j}}}+O(I_{\textrm{i}}^{m}\rho_{\textrm{R}_{\textrm{i}},\textrm{R}_{\textrm{j}}})\right),
		\end{split}
	\end{equation}
	\begin{equation}
		\begin{split}
			\langle \mathrm{S}(2,\textrm{2N-2})|\rho (\mathbf{q})|\mathrm{S}1,2N-1)\rangle]=& \frac{1}{\sqrt{2(1+\mathcal{I}_{N,\mathrm{S}})}}\Bigg[4\rho_{\textrm{L}_{1},\textrm{L}_{1}}I_{{N}}+2\rho_{\textrm{R}_{{N}},\textrm{L}_{1}}+2\rho_{\textrm{L}_{1},\textrm{R}_{{N}}}+2\sum_{\textrm{i}=1}^{{N}-1}(4I_{{N}}\rho_{\textrm{R}_{\textrm{i}},\textrm{R}_{\textrm{i}}}\\
			&-I_{\textrm{i}}(\rho_{\textrm{R}_{\textrm{i}},\textrm{R}_{{N}}}+\rho_{\textrm{R}_{{N}},\textrm{R}_{\textrm{i}}})-2I_{\textrm{i}}I_{{N}}(\rho_{\textrm{R}_{\textrm{i}},\textrm{L}_{1}}+\rho_{\textrm{L}_{1},\textrm{R}_{\textrm{i}}})-I_{\textrm{i}}^{2}(\rho_{\textrm{R}_{{N}},\textrm{L}_{1}}+\rho_{\textrm{L}_{1},\textrm{R}_{{N}}}))\Bigg].
		\end{split}
	\end{equation}
\end{widetext}
Here,$O(I_{\textrm{i}}^{m}\rho_{\textrm{R}_{\textrm{i}},\textrm{R}_{\textrm{j}}})$, $m>2$ are higher-order terms that can be ignored due to the fact $I_{i}\ll 1$.
\begin{figure}
	(a)\includegraphics[scale=0.47]{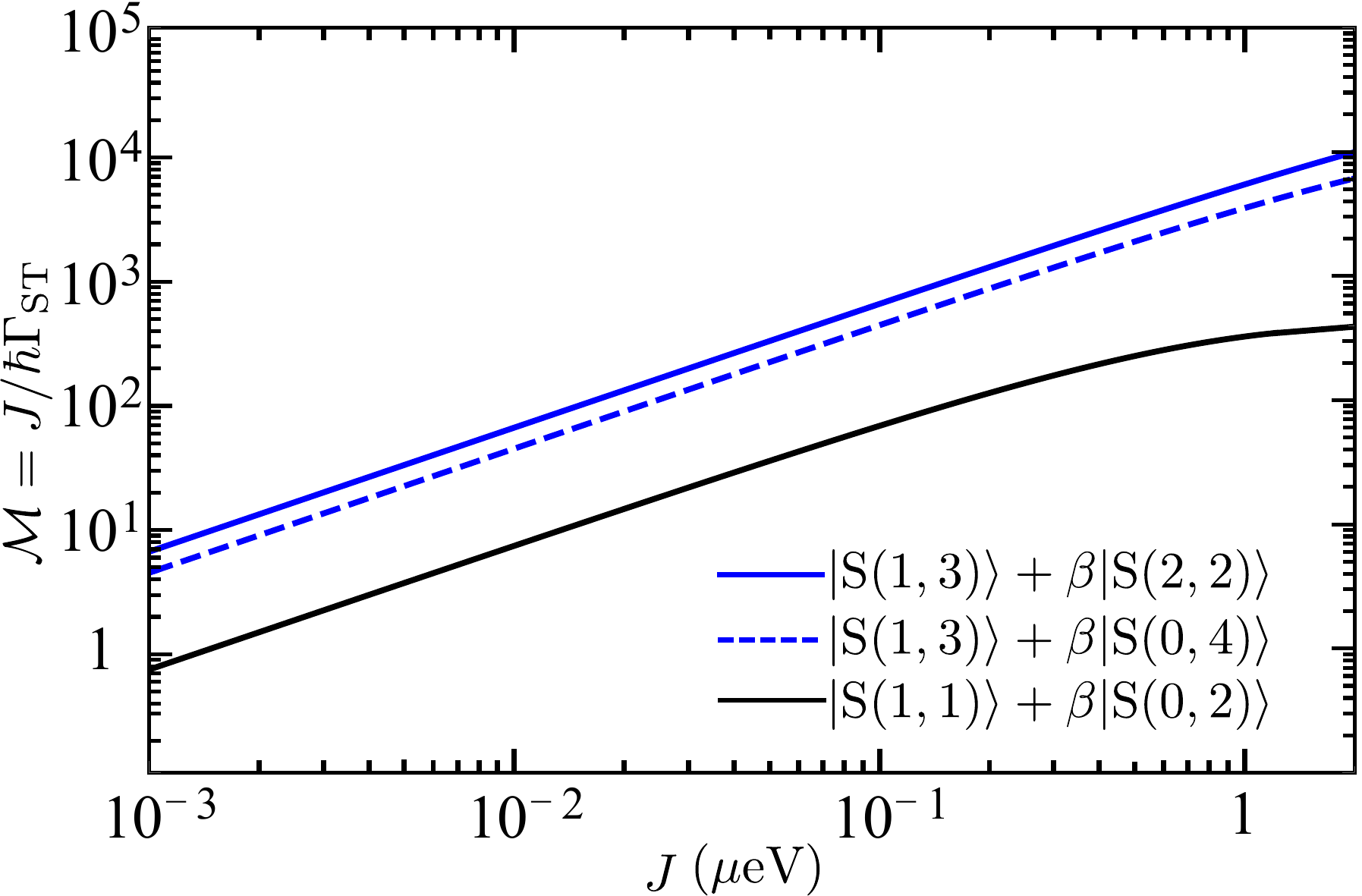}
	(b)\includegraphics[scale=0.466]{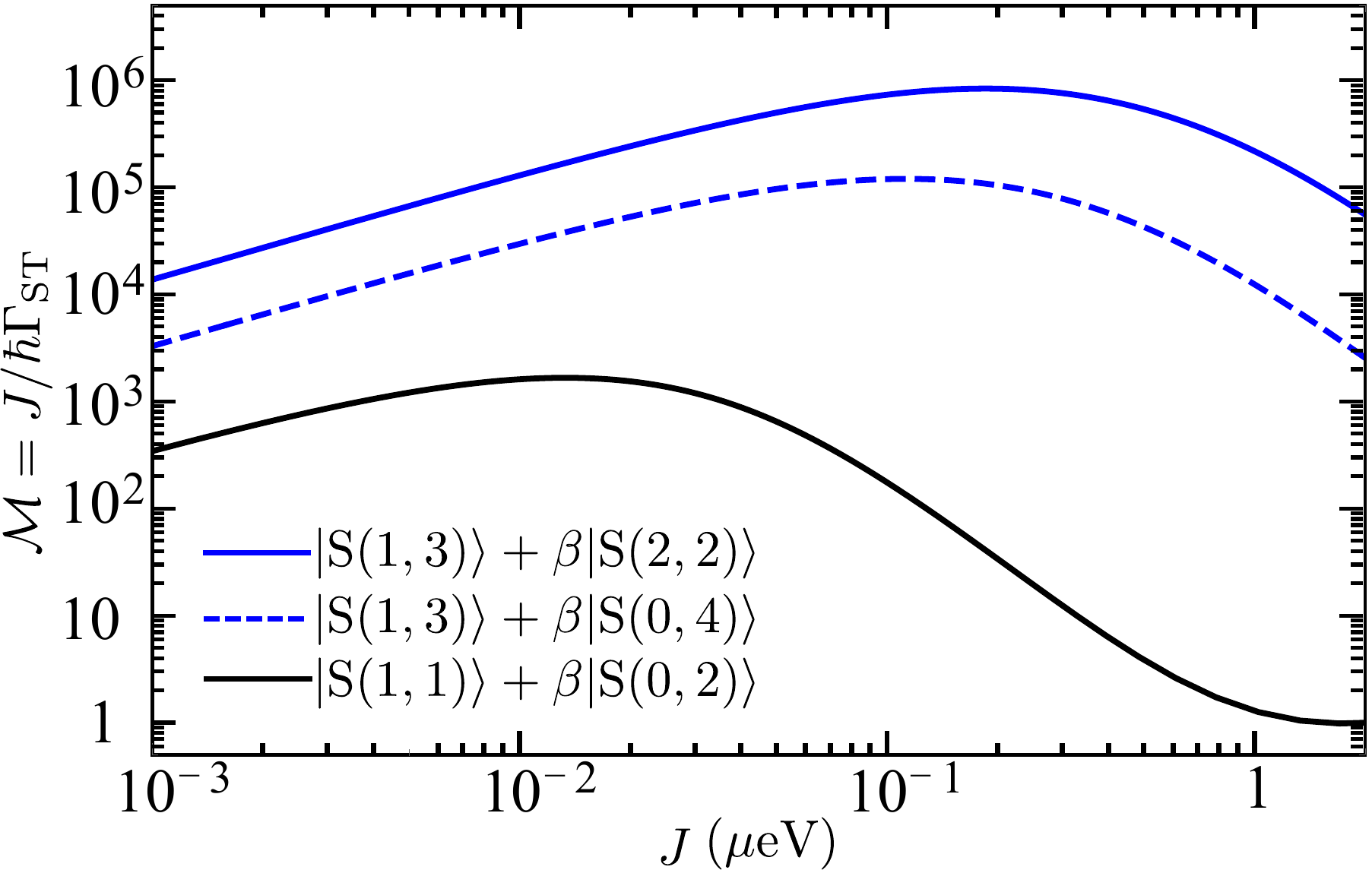}
	(c)\includegraphics[scale=0.636]{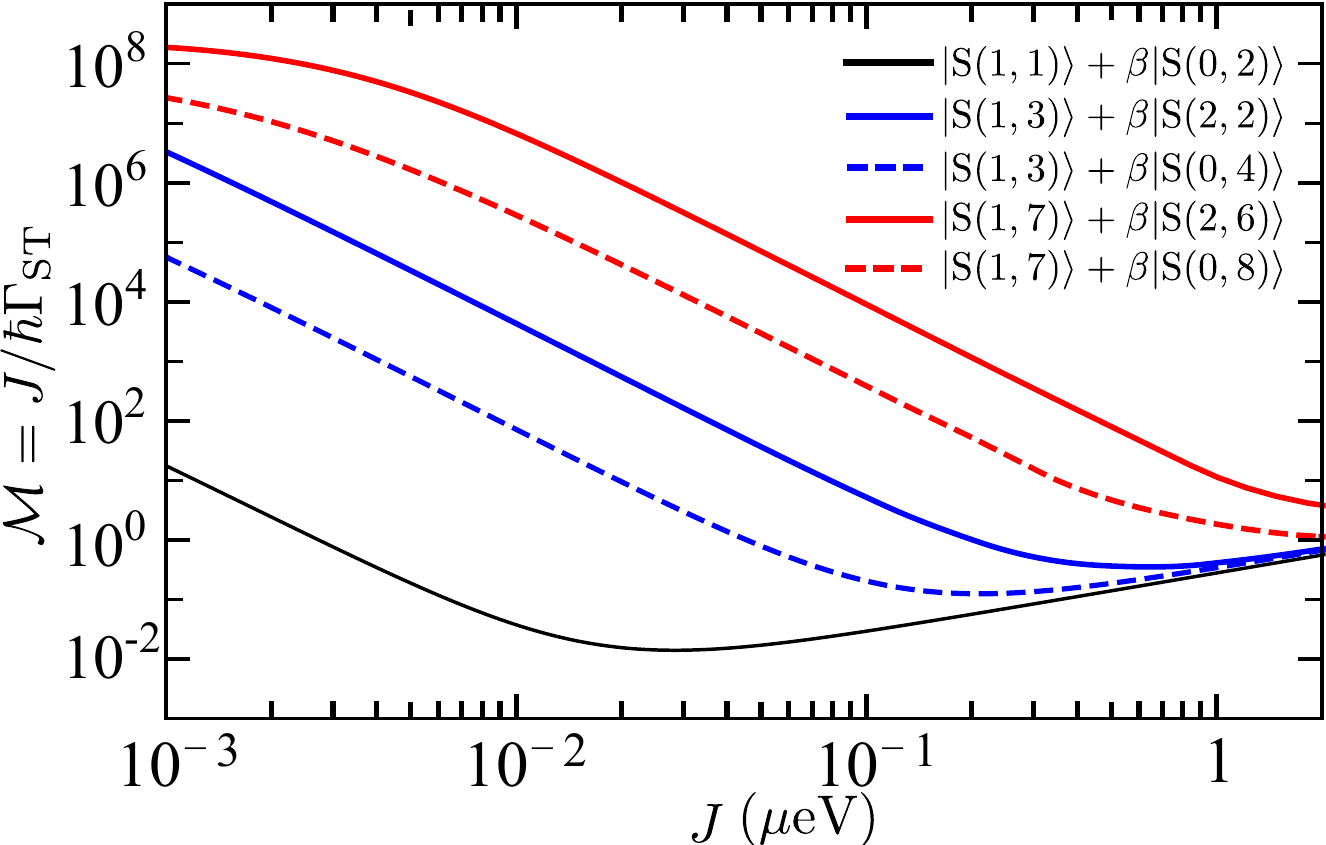}
	\caption{The merit figures v.s.~the exchange interaction calculated for three different sets of parameters. (a) $x_{0}=50  $ nm, $\hbar \omega _{\textrm{L}}=2.838$ meV, $\hbar \omega _{\textrm{R}}=1.419 $ meV. (b) $x_{0}=70  $ nm, $\hbar \omega _{\textrm{L}}=2.838 $ meV, $\hbar \omega _{\textrm{R}}=0.946 $ meV. (c) $x_{0}=80  $ nm, $\hbar \omega _{\textrm{L}}=2.838 $ meV, $\hbar \omega _{\textrm{R}}=1.419 $ meV.}
	\label{fig:Fig6}
\end{figure}

\section{Merit figures of other quantum dot parameters}
\label{Merit figures of other quantum dot parameters}

In Fig.~\ref{fig:Fig6}, we show the merit figures calculated for three sets
of parameters. Fig.~\ref{fig:Fig6}(a) shows a case with a short half dot distance $x_0=50$ nm with a relatively strong confinement strength in the right dot $\hbar\omega_\mathrm{R}=1.419$ meV, while Fig.~\ref{fig:Fig6}(b) shows a case with an intermediate half dot distance $x_0=70$ nm with a relatively weak $\hbar\omega_\mathrm{R}=0.946$ meV, In these cases, the barrier between the two dots is low, rendering the case with $(1,7)$ electron occupancy ill-defined. Therefore only results for $(1,3)$ and $(1,1)$ are shown. We can see that the merit figure associated with $(1,3)$ is clearly higher than those with $(1,1)$, consistent with the findings in the main text. In Fig.~\ref{fig:Fig6}(c), results for all three cases of $(1,7)$, $(1,3)$, and $(1,1)$ are shown. Again, these results are consistent with the main finding that in certain $J$ range, the merit figure for $(1,7)$ is the highest, that for $(1,3)$ is intermediate, and that for $(1,1)$ is the lowest.



\bibliographystyle{apsrev4-1}

\begin{thebibliography}{69}%
\makeatletter
\providecommand \@ifxundefined [1]{%
 \@ifx{#1\undefined}
}%
\providecommand \@ifnum [1]{%
 \ifnum #1\expandafter \@firstoftwo
 \else \expandafter \@secondoftwo
 \fi
}%
\providecommand \@ifx [1]{%
 \ifx #1\expandafter \@firstoftwo
 \else \expandafter \@secondoftwo
 \fi
}%
\providecommand \natexlab [1]{#1}%
\providecommand \enquote  [1]{``#1''}%
\providecommand \bibnamefont  [1]{#1}%
\providecommand \bibfnamefont [1]{#1}%
\providecommand \citenamefont [1]{#1}%
\providecommand \href@noop [0]{\@secondoftwo}%
\providecommand \href [0]{\begingroup \@sanitize@url \@href}%
\providecommand \@href[1]{\@@startlink{#1}\@@href}%
\providecommand \@@href[1]{\endgroup#1\@@endlink}%
\providecommand \@sanitize@url [0]{\catcode `\\12\catcode `\$12\catcode
  `\&12\catcode `\#12\catcode `\^12\catcode `\_12\catcode `\%12\relax}%
\providecommand \@@startlink[1]{}%
\providecommand \@@endlink[0]{}%
\providecommand \url  [0]{\begingroup\@sanitize@url \@url }%
\providecommand \@url [1]{\endgroup\@href {#1}{\urlprefix }}%
\providecommand \urlprefix  [0]{URL }%
\providecommand \Eprint [0]{\href }%
\providecommand \doibase [0]{http://dx.doi.org/}%
\providecommand \selectlanguage [0]{\@gobble}%
\providecommand \bibinfo  [0]{\@secondoftwo}%
\providecommand \bibfield  [0]{\@secondoftwo}%
\providecommand \translation [1]{[#1]}%
\providecommand \BibitemOpen [0]{}%
\providecommand \bibitemStop [0]{}%
\providecommand \bibitemNoStop [0]{.\EOS\space}%
\providecommand \EOS [0]{\spacefactor3000\relax}%
\providecommand \BibitemShut  [1]{\csname bibitem#1\endcsname}%
\let\auto@bib@innerbib\@empty
\bibitem [{\citenamefont {Loss}\ and\ \citenamefont
  {DiVincenzo}(1998)}]{PhysRevA.57.120}%
  \BibitemOpen
  \bibfield  {author} {\bibinfo {author} {\bibfnamefont {D.}~\bibnamefont
  {Loss}}\ and\ \bibinfo {author} {\bibfnamefont {D.~P.}\ \bibnamefont
  {DiVincenzo}},\ }\href {https://link.aps.org/doi/10.1103/PhysRevA.57.120}
  {\bibfield  {journal} {\bibinfo  {journal} {Phys. Rev. A}\ }\textbf {\bibinfo
  {volume} {57}},\ \bibinfo {pages} {120} (\bibinfo {year} {1998})}\BibitemShut
  {NoStop}%
\bibitem [{\citenamefont {Zhang}\ \emph {et~al.}(2018)\citenamefont {Zhang},
  \citenamefont {Li}, \citenamefont {Cao}, \citenamefont {Xiao}, \citenamefont
  {Guo},\ and\ \citenamefont {Guo}}]{10.1093/nsr/nwy153}%
  \BibitemOpen
  \bibfield  {author} {\bibinfo {author} {\bibfnamefont {X.}~\bibnamefont
  {Zhang}}, \bibinfo {author} {\bibfnamefont {H.-O.}\ \bibnamefont {Li}},
  \bibinfo {author} {\bibfnamefont {G.}~\bibnamefont {Cao}}, \bibinfo {author}
  {\bibfnamefont {M.}~\bibnamefont {Xiao}}, \bibinfo {author} {\bibfnamefont
  {G.-C.}\ \bibnamefont {Guo}}, \ and\ \bibinfo {author} {\bibfnamefont
  {G.-P.}\ \bibnamefont {Guo}},\ }\href {https://doi.org/10.1093/nsr/nwy153}
  {\bibfield  {journal} {\bibinfo  {journal} {Natl. Sci. Rev.}\ }\textbf
  {\bibinfo {volume} {6}},\ \bibinfo {pages} {32} (\bibinfo {year}
  {2018})}\BibitemShut {NoStop}%
\bibitem [{\citenamefont {Yang}\ \emph {et~al.}(2018)\citenamefont {Yang},
  \citenamefont {Chan},\ and\ \citenamefont {Wang}}]{PhysRevA.98.032334}%
  \BibitemOpen
  \bibfield  {author} {\bibinfo {author} {\bibfnamefont {X.-C.}\ \bibnamefont
  {Yang}}, \bibinfo {author} {\bibfnamefont {G.~X.}\ \bibnamefont {Chan}}, \
  and\ \bibinfo {author} {\bibfnamefont {X.}~\bibnamefont {Wang}},\ }\href
  {https://link.aps.org/doi/10.1103/PhysRevA.98.032334} {\bibfield  {journal}
  {\bibinfo  {journal} {Phys. Rev. A}\ }\textbf {\bibinfo {volume} {98}},\
  \bibinfo {pages} {032334} (\bibinfo {year} {2018})}\BibitemShut {NoStop}%
\bibitem [{\citenamefont {Burkard}\ \emph {et~al.}(1999)\citenamefont
  {Burkard}, \citenamefont {Loss},\ and\ \citenamefont
  {DiVincenzo}}]{PhysRevB.59.2070}%
  \BibitemOpen
  \bibfield  {author} {\bibinfo {author} {\bibfnamefont {G.}~\bibnamefont
  {Burkard}}, \bibinfo {author} {\bibfnamefont {D.}~\bibnamefont {Loss}}, \
  and\ \bibinfo {author} {\bibfnamefont {D.~P.}\ \bibnamefont {DiVincenzo}},\
  }\href {\doibase 10.1103/PhysRevB.59.2070} {\bibfield  {journal} {\bibinfo
  {journal} {Phys. Rev. B}\ }\textbf {\bibinfo {volume} {59}},\ \bibinfo
  {pages} {2070} (\bibinfo {year} {1999})}\BibitemShut {NoStop}%
\bibitem [{\citenamefont {Sala}\ \emph {et~al.}(2020)\citenamefont {Sala},
  \citenamefont {Qvist},\ and\ \citenamefont
  {Danon}}]{PhysRevResearch.2.012062}%
  \BibitemOpen
  \bibfield  {author} {\bibinfo {author} {\bibfnamefont {A.}~\bibnamefont
  {Sala}}, \bibinfo {author} {\bibfnamefont {J.~H.}\ \bibnamefont {Qvist}}, \
  and\ \bibinfo {author} {\bibfnamefont {J.}~\bibnamefont {Danon}},\ }\href
  {\doibase 10.1103/PhysRevResearch.2.012062} {\bibfield  {journal} {\bibinfo
  {journal} {Phys. Rev. Research}\ }\textbf {\bibinfo {volume} {2}},\ \bibinfo
  {pages} {012062} (\bibinfo {year} {2020})}\BibitemShut {NoStop}%
\bibitem [{\citenamefont {Petersson}\ \emph {et~al.}(2010)\citenamefont
  {Petersson}, \citenamefont {Petta}, \citenamefont {Lu},\ and\ \citenamefont
  {Gossard}}]{PhysRevLett.105.246804}%
  \BibitemOpen
  \bibfield  {author} {\bibinfo {author} {\bibfnamefont {K.~D.}\ \bibnamefont
  {Petersson}}, \bibinfo {author} {\bibfnamefont {J.~R.}\ \bibnamefont
  {Petta}}, \bibinfo {author} {\bibfnamefont {H.}~\bibnamefont {Lu}}, \ and\
  \bibinfo {author} {\bibfnamefont {A.~C.}\ \bibnamefont {Gossard}},\ }\href
  {\doibase 10.1103/PhysRevLett.105.246804} {\bibfield  {journal} {\bibinfo
  {journal} {Phys. Rev. Lett.}\ }\textbf {\bibinfo {volume} {105}},\ \bibinfo
  {pages} {246804} (\bibinfo {year} {2010})}\BibitemShut {NoStop}%
\bibitem [{\citenamefont {Hensgens}\ \emph {et~al.}(2017)\citenamefont
  {Hensgens}, \citenamefont {Fujita}, \citenamefont {Janssen}, \citenamefont
  {Li}, \citenamefont {Van~Diepen}, \citenamefont {Reichl}, \citenamefont
  {Wegscheider}, \citenamefont {Das~Sarma},\ and\ \citenamefont
  {Vandersypen}}]{10.1038/nature23022}%
  \BibitemOpen
  \bibfield  {author} {\bibinfo {author} {\bibfnamefont {T.}~\bibnamefont
  {Hensgens}}, \bibinfo {author} {\bibfnamefont {T.}~\bibnamefont {Fujita}},
  \bibinfo {author} {\bibfnamefont {L.}~\bibnamefont {Janssen}}, \bibinfo
  {author} {\bibfnamefont {X.}~\bibnamefont {Li}}, \bibinfo {author}
  {\bibfnamefont {C.~J.}\ \bibnamefont {Van~Diepen}}, \bibinfo {author}
  {\bibfnamefont {C.}~\bibnamefont {Reichl}}, \bibinfo {author} {\bibfnamefont
  {W.}~\bibnamefont {Wegscheider}}, \bibinfo {author} {\bibfnamefont
  {S.}~\bibnamefont {Das~Sarma}}, \ and\ \bibinfo {author} {\bibfnamefont
  {L.~M.~K.}\ \bibnamefont {Vandersypen}},\ }\href {\doibase
  10.1038/nature23022} {\bibfield  {journal} {\bibinfo  {journal} {Nature}\
  }\textbf {\bibinfo {volume} {548}},\ \bibinfo {pages} {70} (\bibinfo {year}
  {2017})}\BibitemShut {NoStop}%
\bibitem [{\citenamefont {Wolfe}\ \emph {et~al.}(2017)\citenamefont {Wolfe},
  \citenamefont {Calderon-Vargas},\ and\ \citenamefont
  {Kestner}}]{PhysRevB.96.201307}%
  \BibitemOpen
  \bibfield  {author} {\bibinfo {author} {\bibfnamefont {M.~A.}\ \bibnamefont
  {Wolfe}}, \bibinfo {author} {\bibfnamefont {F.~A.}\ \bibnamefont
  {Calderon-Vargas}}, \ and\ \bibinfo {author} {\bibfnamefont {J.~P.}\
  \bibnamefont {Kestner}},\ }\href {\doibase 10.1103/PhysRevB.96.201307}
  {\bibfield  {journal} {\bibinfo  {journal} {Phys. Rev. B}\ }\textbf {\bibinfo
  {volume} {96}},\ \bibinfo {pages} {201307} (\bibinfo {year}
  {2017})}\BibitemShut {NoStop}%
\bibitem [{\citenamefont {Baruffa}\ \emph {et~al.}(2010)\citenamefont
  {Baruffa}, \citenamefont {Stano},\ and\ \citenamefont
  {Fabian}}]{PhysRevB.82.045311}%
  \BibitemOpen
  \bibfield  {author} {\bibinfo {author} {\bibfnamefont {F.}~\bibnamefont
  {Baruffa}}, \bibinfo {author} {\bibfnamefont {P.}~\bibnamefont {Stano}}, \
  and\ \bibinfo {author} {\bibfnamefont {J.}~\bibnamefont {Fabian}},\ }\href
  {\doibase 10.1103/PhysRevB.82.045311} {\bibfield  {journal} {\bibinfo
  {journal} {Phys. Rev. B}\ }\textbf {\bibinfo {volume} {82}},\ \bibinfo
  {pages} {045311} (\bibinfo {year} {2010})}\BibitemShut {NoStop}%
\bibitem [{\citenamefont {Fedele}\ \emph {et~al.}(2021)\citenamefont {Fedele},
  \citenamefont {Chatterjee}, \citenamefont {Fallahi}, \citenamefont {Gardner},
  \citenamefont {Manfra},\ and\ \citenamefont
  {Kuemmeth}}]{PRXQuantum.2.040306}%
  \BibitemOpen
  \bibfield  {author} {\bibinfo {author} {\bibfnamefont {F.}~\bibnamefont
  {Fedele}}, \bibinfo {author} {\bibfnamefont {A.}~\bibnamefont {Chatterjee}},
  \bibinfo {author} {\bibfnamefont {S.}~\bibnamefont {Fallahi}}, \bibinfo
  {author} {\bibfnamefont {G.~C.}\ \bibnamefont {Gardner}}, \bibinfo {author}
  {\bibfnamefont {M.~J.}\ \bibnamefont {Manfra}}, \ and\ \bibinfo {author}
  {\bibfnamefont {F.}~\bibnamefont {Kuemmeth}},\ }\href {\doibase
  10.1103/PRXQuantum.2.040306} {\bibfield  {journal} {\bibinfo  {journal} {PRX
  Quantum}\ }\textbf {\bibinfo {volume} {2}},\ \bibinfo {pages} {040306}
  (\bibinfo {year} {2021})}\BibitemShut {NoStop}%
\bibitem [{\citenamefont {Feng}\ \emph {et~al.}(2021)\citenamefont {Feng},
  \citenamefont {Zaw},\ and\ \citenamefont {Koh}}]{10.1038/s41534-021-00449-4}%
  \BibitemOpen
  \bibfield  {author} {\bibinfo {author} {\bibfnamefont {M.}~\bibnamefont
  {Feng}}, \bibinfo {author} {\bibfnamefont {L.~H.}\ \bibnamefont {Zaw}}, \
  and\ \bibinfo {author} {\bibfnamefont {T.~S.}\ \bibnamefont {Koh}},\ }\href
  {http://dx.doi.org/10.1038/s41534-021-00449-4} {\bibfield  {journal}
  {\bibinfo  {journal} {npj Quantum Inf.}\ }\textbf {\bibinfo {volume} {7}},\
  \bibinfo {pages} {112} (\bibinfo {year} {2021})}\BibitemShut {NoStop}%
\bibitem [{\citenamefont {Harvey-Collard}\ \emph {et~al.}(2019)\citenamefont
  {Harvey-Collard}, \citenamefont {Jacobson}, \citenamefont {Bureau-Oxton},
  \citenamefont {Jock}, \citenamefont {Srinivasa}, \citenamefont {Mounce},
  \citenamefont {Ward}, \citenamefont {Anderson}, \citenamefont {Manginell},
  \citenamefont {Wendt}, \citenamefont {Pluym}, \citenamefont {Lilly},
  \citenamefont {Luhman}, \citenamefont {Pioro-Ladri\`ere},\ and\ \citenamefont
  {Carroll}}]{PhysRevLett.122.217702}%
  \BibitemOpen
  \bibfield  {author} {\bibinfo {author} {\bibfnamefont {P.}~\bibnamefont
  {Harvey-Collard}}, \bibinfo {author} {\bibfnamefont {N.~T.}\ \bibnamefont
  {Jacobson}}, \bibinfo {author} {\bibfnamefont {C.}~\bibnamefont
  {Bureau-Oxton}}, \bibinfo {author} {\bibfnamefont {R.~M.}\ \bibnamefont
  {Jock}}, \bibinfo {author} {\bibfnamefont {V.}~\bibnamefont {Srinivasa}},
  \bibinfo {author} {\bibfnamefont {A.~M.}\ \bibnamefont {Mounce}}, \bibinfo
  {author} {\bibfnamefont {D.~R.}\ \bibnamefont {Ward}}, \bibinfo {author}
  {\bibfnamefont {J.~M.}\ \bibnamefont {Anderson}}, \bibinfo {author}
  {\bibfnamefont {R.~P.}\ \bibnamefont {Manginell}}, \bibinfo {author}
  {\bibfnamefont {J.~R.}\ \bibnamefont {Wendt}}, \bibinfo {author}
  {\bibfnamefont {T.}~\bibnamefont {Pluym}}, \bibinfo {author} {\bibfnamefont
  {M.~P.}\ \bibnamefont {Lilly}}, \bibinfo {author} {\bibfnamefont {D.~R.}\
  \bibnamefont {Luhman}}, \bibinfo {author} {\bibfnamefont {M.}~\bibnamefont
  {Pioro-Ladri\`ere}}, \ and\ \bibinfo {author} {\bibfnamefont {M.~S.}\
  \bibnamefont {Carroll}},\ }\href {\doibase 10.1103/PhysRevLett.122.217702}
  {\bibfield  {journal} {\bibinfo  {journal} {Phys. Rev. Lett.}\ }\textbf
  {\bibinfo {volume} {122}},\ \bibinfo {pages} {217702} (\bibinfo {year}
  {2019})}\BibitemShut {NoStop}%
\bibitem [{\citenamefont {Climente}\ \emph {et~al.}(2007)\citenamefont
  {Climente}, \citenamefont {Bertoni}, \citenamefont {Goldoni}, \citenamefont
  {Rontani},\ and\ \citenamefont {Molinari}}]{PhysRevB.75.081303}%
  \BibitemOpen
  \bibfield  {author} {\bibinfo {author} {\bibfnamefont {J.~I.}\ \bibnamefont
  {Climente}}, \bibinfo {author} {\bibfnamefont {A.}~\bibnamefont {Bertoni}},
  \bibinfo {author} {\bibfnamefont {G.}~\bibnamefont {Goldoni}}, \bibinfo
  {author} {\bibfnamefont {M.}~\bibnamefont {Rontani}}, \ and\ \bibinfo
  {author} {\bibfnamefont {E.}~\bibnamefont {Molinari}},\ }\href {\doibase
  10.1103/PhysRevB.75.081303} {\bibfield  {journal} {\bibinfo  {journal} {Phys.
  Rev. B}\ }\textbf {\bibinfo {volume} {75}},\ \bibinfo {pages} {081303}
  (\bibinfo {year} {2007})}\BibitemShut {NoStop}%
\bibitem [{\citenamefont {Deng}\ and\ \citenamefont
  {Barnes}(2020)}]{PhysRevB.102.035427}%
  \BibitemOpen
  \bibfield  {author} {\bibinfo {author} {\bibfnamefont {K.}~\bibnamefont
  {Deng}}\ and\ \bibinfo {author} {\bibfnamefont {E.}~\bibnamefont {Barnes}},\
  }\href {\doibase 10.1103/PhysRevB.102.035427} {\bibfield  {journal} {\bibinfo
   {journal} {Phys. Rev. B}\ }\textbf {\bibinfo {volume} {102}},\ \bibinfo
  {pages} {035427} (\bibinfo {year} {2020})}\BibitemShut {NoStop}%
\bibitem [{\citenamefont {Barnes}\ \emph {et~al.}(2011)\citenamefont {Barnes},
  \citenamefont {Kestner}, \citenamefont {Nguyen},\ and\ \citenamefont
  {Das~Sarma}}]{PhysRevB.84.235309}%
  \BibitemOpen
  \bibfield  {author} {\bibinfo {author} {\bibfnamefont {E.}~\bibnamefont
  {Barnes}}, \bibinfo {author} {\bibfnamefont {J.~P.}\ \bibnamefont {Kestner}},
  \bibinfo {author} {\bibfnamefont {N.~T.~T.}\ \bibnamefont {Nguyen}}, \ and\
  \bibinfo {author} {\bibfnamefont {S.}~\bibnamefont {Das~Sarma}},\ }\href
  {\doibase 10.1103/PhysRevB.84.235309} {\bibfield  {journal} {\bibinfo
  {journal} {Phys. Rev. B}\ }\textbf {\bibinfo {volume} {84}},\ \bibinfo
  {pages} {235309} (\bibinfo {year} {2011})}\BibitemShut {NoStop}%
\bibitem [{\citenamefont {Stepanenko}\ \emph {et~al.}(2003)\citenamefont
  {Stepanenko}, \citenamefont {Bonesteel}, \citenamefont {DiVincenzo},
  \citenamefont {Burkard},\ and\ \citenamefont {Loss}}]{PhysRevB.68.115306}%
  \BibitemOpen
  \bibfield  {author} {\bibinfo {author} {\bibfnamefont {D.}~\bibnamefont
  {Stepanenko}}, \bibinfo {author} {\bibfnamefont {N.~E.}\ \bibnamefont
  {Bonesteel}}, \bibinfo {author} {\bibfnamefont {D.~P.}\ \bibnamefont
  {DiVincenzo}}, \bibinfo {author} {\bibfnamefont {G.}~\bibnamefont {Burkard}},
  \ and\ \bibinfo {author} {\bibfnamefont {D.}~\bibnamefont {Loss}},\ }\href
  {\doibase 10.1103/PhysRevB.68.115306} {\bibfield  {journal} {\bibinfo
  {journal} {Phys. Rev. B}\ }\textbf {\bibinfo {volume} {68}},\ \bibinfo
  {pages} {115306} (\bibinfo {year} {2003})}\BibitemShut {NoStop}%
\bibitem [{\citenamefont {Stepanenko}\ \emph {et~al.}(2012)\citenamefont
  {Stepanenko}, \citenamefont {Rudner}, \citenamefont {Halperin},\ and\
  \citenamefont {Loss}}]{PhysRevB.85.075416}%
  \BibitemOpen
  \bibfield  {author} {\bibinfo {author} {\bibfnamefont {D.}~\bibnamefont
  {Stepanenko}}, \bibinfo {author} {\bibfnamefont {M.}~\bibnamefont {Rudner}},
  \bibinfo {author} {\bibfnamefont {B.~I.}\ \bibnamefont {Halperin}}, \ and\
  \bibinfo {author} {\bibfnamefont {D.}~\bibnamefont {Loss}},\ }\href {\doibase
  10.1103/PhysRevB.85.075416} {\bibfield  {journal} {\bibinfo  {journal} {Phys.
  Rev. B}\ }\textbf {\bibinfo {volume} {85}},\ \bibinfo {pages} {075416}
  (\bibinfo {year} {2012})}\BibitemShut {NoStop}%
\bibitem [{\citenamefont {Chan}\ and\ \citenamefont
  {Wang}(2019)}]{10.1002/qute.201900072}%
  \BibitemOpen
  \bibfield  {author} {\bibinfo {author} {\bibfnamefont {G.~X.}\ \bibnamefont
  {Chan}}\ and\ \bibinfo {author} {\bibfnamefont {X.}~\bibnamefont {Wang}},\
  }\href {http://dx.doi.org/10.1002/qute.201900072} {\bibfield  {journal}
  {\bibinfo  {journal} {Adv. Quantum Technol.}\ }\textbf {\bibinfo {volume}
  {2}},\ \bibinfo {pages} {1900072} (\bibinfo {year} {2019})}\BibitemShut
  {NoStop}%
\bibitem [{\citenamefont {Chan}\ \emph {et~al.}(2021)\citenamefont {Chan},
  \citenamefont {Kestner},\ and\ \citenamefont {Wang}}]{PhysRevB.103.L161409}%
  \BibitemOpen
  \bibfield  {author} {\bibinfo {author} {\bibfnamefont {G.~X.}\ \bibnamefont
  {Chan}}, \bibinfo {author} {\bibfnamefont {J.~P.}\ \bibnamefont {Kestner}}, \
  and\ \bibinfo {author} {\bibfnamefont {X.}~\bibnamefont {Wang}},\ }\href
  {https://link.aps.org/doi/10.1103/PhysRevB.103.L161409} {\bibfield  {journal}
  {\bibinfo  {journal} {Phys. Rev. B}\ }\textbf {\bibinfo {volume} {103}},\
  \bibinfo {pages} {L161409} (\bibinfo {year} {2021})}\BibitemShut {NoStop}%
\bibitem [{\citenamefont {Szabo}\ \emph {et~al.}(1982)\citenamefont {Szabo},
  \citenamefont {Szab{\'o}},\ and\ \citenamefont {Ostlund}}]{szabo1982modern}%
  \BibitemOpen
  \bibfield  {author} {\bibinfo {author} {\bibfnamefont {A.}~\bibnamefont
  {Szabo}}, \bibinfo {author} {\bibfnamefont {A.}~\bibnamefont {Szab{\'o}}}, \
  and\ \bibinfo {author} {\bibfnamefont {N.}~\bibnamefont {Ostlund}},\ }\href
  {https://books.google.com.hk/books?id=1ky8QgAACAAJ} {\emph {\bibinfo {title}
  {Modern Quantum Chemistry: Introduction to Advanced Electronic Structure
  Theory}}}\ (\bibinfo  {publisher} {Macmillan},\ \bibinfo {year}
  {1982})\BibitemShut {NoStop}%
\bibitem [{\citenamefont {Maune}\ \emph {et~al.}(2012)\citenamefont {Maune},
  \citenamefont {Borselli}, \citenamefont {Huang}, \citenamefont {Ladd},
  \citenamefont {Deelman}, \citenamefont {Holabird}, \citenamefont {Kiselev},
  \citenamefont {Alvarado-Rodriguez}, \citenamefont {Ross}, \citenamefont
  {Schmitz} \emph {et~al.}}]{maune2012coherent}%
  \BibitemOpen
  \bibfield  {author} {\bibinfo {author} {\bibfnamefont {B.~M.}\ \bibnamefont
  {Maune}}, \bibinfo {author} {\bibfnamefont {M.~G.}\ \bibnamefont {Borselli}},
  \bibinfo {author} {\bibfnamefont {B.}~\bibnamefont {Huang}}, \bibinfo
  {author} {\bibfnamefont {T.~D.}\ \bibnamefont {Ladd}}, \bibinfo {author}
  {\bibfnamefont {P.~W.}\ \bibnamefont {Deelman}}, \bibinfo {author}
  {\bibfnamefont {K.~S.}\ \bibnamefont {Holabird}}, \bibinfo {author}
  {\bibfnamefont {A.~A.}\ \bibnamefont {Kiselev}}, \bibinfo {author}
  {\bibfnamefont {I.}~\bibnamefont {Alvarado-Rodriguez}}, \bibinfo {author}
  {\bibfnamefont {R.~S.}\ \bibnamefont {Ross}}, \bibinfo {author}
  {\bibfnamefont {A.~E.}\ \bibnamefont {Schmitz}},  \emph {et~al.},\ }\href
  {https://doi.org/10.1038/nature10707} {\bibfield  {journal} {\bibinfo
  {journal} {Nature}\ }\textbf {\bibinfo {volume} {481}},\ \bibinfo {pages}
  {344} (\bibinfo {year} {2012})}\BibitemShut {NoStop}%
\bibitem [{\citenamefont {van Diepen}\ \emph {et~al.}(2021)\citenamefont {van
  Diepen}, \citenamefont {Hsiao}, \citenamefont {Mukhopadhyay}, \citenamefont
  {Reichl}, \citenamefont {Wegscheider},\ and\ \citenamefont
  {Vandersypen}}]{PhysRevX.11.041025}%
  \BibitemOpen
  \bibfield  {author} {\bibinfo {author} {\bibfnamefont {C.~J.}\ \bibnamefont
  {van Diepen}}, \bibinfo {author} {\bibfnamefont {T.-K.}\ \bibnamefont
  {Hsiao}}, \bibinfo {author} {\bibfnamefont {U.}~\bibnamefont {Mukhopadhyay}},
  \bibinfo {author} {\bibfnamefont {C.}~\bibnamefont {Reichl}}, \bibinfo
  {author} {\bibfnamefont {W.}~\bibnamefont {Wegscheider}}, \ and\ \bibinfo
  {author} {\bibfnamefont {L.~M.~K.}\ \bibnamefont {Vandersypen}},\ }\href
  {\doibase 10.1103/PhysRevX.11.041025} {\bibfield  {journal} {\bibinfo
  {journal} {Phys. Rev. X}\ }\textbf {\bibinfo {volume} {11}},\ \bibinfo
  {pages} {041025} (\bibinfo {year} {2021})}\BibitemShut {NoStop}%
\bibitem [{\citenamefont {Zajac}\ \emph {et~al.}(2018)\citenamefont {Zajac},
  \citenamefont {Sigillito}, \citenamefont {Russ}, \citenamefont {Borjans},
  \citenamefont {Taylor}, \citenamefont {Burkard},\ and\ \citenamefont
  {Petta}}]{doi:10.1126/science.aao5965}%
  \BibitemOpen
  \bibfield  {author} {\bibinfo {author} {\bibfnamefont {D.~M.}\ \bibnamefont
  {Zajac}}, \bibinfo {author} {\bibfnamefont {A.~J.}\ \bibnamefont
  {Sigillito}}, \bibinfo {author} {\bibfnamefont {M.}~\bibnamefont {Russ}},
  \bibinfo {author} {\bibfnamefont {F.}~\bibnamefont {Borjans}}, \bibinfo
  {author} {\bibfnamefont {J.~M.}\ \bibnamefont {Taylor}}, \bibinfo {author}
  {\bibfnamefont {G.}~\bibnamefont {Burkard}}, \ and\ \bibinfo {author}
  {\bibfnamefont {J.~R.}\ \bibnamefont {Petta}},\ }\href {\doibase
  10.1126/science.aao5965} {\bibfield  {journal} {\bibinfo  {journal}
  {Science}\ }\textbf {\bibinfo {volume} {359}},\ \bibinfo {pages} {439}
  (\bibinfo {year} {2018})}\BibitemShut {NoStop}%
\bibitem [{\citenamefont {Cerfontaine}\ \emph {et~al.}(2020)\citenamefont
  {Cerfontaine}, \citenamefont {Botzem}, \citenamefont {Ritzmann},
  \citenamefont {Humpohl}, \citenamefont {Ludwig}, \citenamefont {Schuh},
  \citenamefont {Bougeard}, \citenamefont {Wieck},\ and\ \citenamefont
  {Bluhm}}]{10.1038/s41467-020-17865-3}%
  \BibitemOpen
  \bibfield  {author} {\bibinfo {author} {\bibfnamefont {P.}~\bibnamefont
  {Cerfontaine}}, \bibinfo {author} {\bibfnamefont {T.}~\bibnamefont {Botzem}},
  \bibinfo {author} {\bibfnamefont {J.}~\bibnamefont {Ritzmann}}, \bibinfo
  {author} {\bibfnamefont {S.~S.}\ \bibnamefont {Humpohl}}, \bibinfo {author}
  {\bibfnamefont {A.}~\bibnamefont {Ludwig}}, \bibinfo {author} {\bibfnamefont
  {D.}~\bibnamefont {Schuh}}, \bibinfo {author} {\bibfnamefont
  {D.}~\bibnamefont {Bougeard}}, \bibinfo {author} {\bibfnamefont {A.~D.}\
  \bibnamefont {Wieck}}, \ and\ \bibinfo {author} {\bibfnamefont
  {H.}~\bibnamefont {Bluhm}},\ }\href
  {http://dx.doi.org/10.1038/s41467-020-17865-3} {\bibfield  {journal}
  {\bibinfo  {journal} {Nat. Commun.}\ }\textbf {\bibinfo {volume} {11}},\
  \bibinfo {pages} {4144} (\bibinfo {year} {2020})}\BibitemShut {NoStop}%
\bibitem [{\citenamefont {Dhayal}\ \emph {et~al.}(2014)\citenamefont {Dhayal},
  \citenamefont {Ramaniah}, \citenamefont {Ruda},\ and\ \citenamefont
  {Nair}}]{del1999electronic}%
  \BibitemOpen
  \bibfield  {author} {\bibinfo {author} {\bibfnamefont {S.~S.}\ \bibnamefont
  {Dhayal}}, \bibinfo {author} {\bibfnamefont {L.~M.}\ \bibnamefont
  {Ramaniah}}, \bibinfo {author} {\bibfnamefont {H.~E.}\ \bibnamefont {Ruda}},
  \ and\ \bibinfo {author} {\bibfnamefont {S.~V.}\ \bibnamefont {Nair}},\
  }\href {https://aip.scitation.org/doi/10.1063/1.4901923} {\bibfield
  {journal} {\bibinfo  {journal} {J. Chem. Phys.}\ }\textbf {\bibinfo {volume}
  {141}},\ \bibinfo {pages} {204702} (\bibinfo {year} {2014})}\BibitemShut
  {NoStop}%
\bibitem [{\citenamefont {Malinowski}\ \emph {et~al.}(2019)\citenamefont
  {Malinowski}, \citenamefont {Martins}, \citenamefont {Smith}, \citenamefont
  {Bartlett}, \citenamefont {Doherty}, \citenamefont {Nissen}, \citenamefont
  {Fallahi}, \citenamefont {Gardner}, \citenamefont {Manfra}, \citenamefont
  {Marcus},\ and\ \citenamefont {Kuemmeth}}]{10.1038/s41467-019-09194-x}%
  \BibitemOpen
  \bibfield  {author} {\bibinfo {author} {\bibfnamefont {F.~K.}\ \bibnamefont
  {Malinowski}}, \bibinfo {author} {\bibfnamefont {F.}~\bibnamefont {Martins}},
  \bibinfo {author} {\bibfnamefont {T.~B.}\ \bibnamefont {Smith}}, \bibinfo
  {author} {\bibfnamefont {S.~D.}\ \bibnamefont {Bartlett}}, \bibinfo {author}
  {\bibfnamefont {A.~C.}\ \bibnamefont {Doherty}}, \bibinfo {author}
  {\bibfnamefont {P.~D.}\ \bibnamefont {Nissen}}, \bibinfo {author}
  {\bibfnamefont {S.}~\bibnamefont {Fallahi}}, \bibinfo {author} {\bibfnamefont
  {G.~C.}\ \bibnamefont {Gardner}}, \bibinfo {author} {\bibfnamefont {M.~J.}\
  \bibnamefont {Manfra}}, \bibinfo {author} {\bibfnamefont {C.~M.}\
  \bibnamefont {Marcus}}, \ and\ \bibinfo {author} {\bibfnamefont
  {F.}~\bibnamefont {Kuemmeth}},\ }\href
  {http://dx.doi.org/10.1038/s41467-019-09194-x} {\bibfield  {journal}
  {\bibinfo  {journal} {Nat. Commun.}\ }\textbf {\bibinfo {volume} {10}},\
  \bibinfo {pages} {1196} (\bibinfo {year} {2019})}\BibitemShut {NoStop}%
\bibitem [{\citenamefont {Mills}\ \emph {et~al.}(2021)\citenamefont {Mills},
  \citenamefont {Guinn}, \citenamefont {Gullans}, \citenamefont {Sigillito},
  \citenamefont {Feldman}, \citenamefont {Nielsen},\ and\ \citenamefont
  {Petta}}]{mills2021two}%
  \BibitemOpen
  \bibfield  {author} {\bibinfo {author} {\bibfnamefont {A.}~\bibnamefont
  {Mills}}, \bibinfo {author} {\bibfnamefont {C.}~\bibnamefont {Guinn}},
  \bibinfo {author} {\bibfnamefont {M.}~\bibnamefont {Gullans}}, \bibinfo
  {author} {\bibfnamefont {A.}~\bibnamefont {Sigillito}}, \bibinfo {author}
  {\bibfnamefont {M.}~\bibnamefont {Feldman}}, \bibinfo {author} {\bibfnamefont
  {E.}~\bibnamefont {Nielsen}}, \ and\ \bibinfo {author} {\bibfnamefont
  {J.}~\bibnamefont {Petta}},\ }\href {https://arxiv.org/abs/2111.11937}
  {\bibfield  {journal} {\bibinfo  {journal} {arXiv preprint arXiv:2111.11937}\
  } (\bibinfo {year} {2021})}\BibitemShut {NoStop}%
\bibitem [{\citenamefont {Wang}\ \emph {et~al.}(2022)\citenamefont {Wang},
  \citenamefont {Wang}, \citenamefont {Chen}, \citenamefont {Zhang},
  \citenamefont {Li}, \citenamefont {Nan}, \citenamefont {Wang}, \citenamefont
  {Yang}, \citenamefont {Laref},\ and\ \citenamefont
  {Xiong}}]{PhysRevB.105.075430}%
  \BibitemOpen
  \bibfield  {author} {\bibinfo {author} {\bibfnamefont {P.-C.}\ \bibnamefont
  {Wang}}, \bibinfo {author} {\bibfnamefont {Y.-H.}\ \bibnamefont {Wang}},
  \bibinfo {author} {\bibfnamefont {C.}~\bibnamefont {Chen}}, \bibinfo {author}
  {\bibfnamefont {J.}~\bibnamefont {Zhang}}, \bibinfo {author} {\bibfnamefont
  {W.}~\bibnamefont {Li}}, \bibinfo {author} {\bibfnamefont {N.}~\bibnamefont
  {Nan}}, \bibinfo {author} {\bibfnamefont {J.-N.}\ \bibnamefont {Wang}},
  \bibinfo {author} {\bibfnamefont {J.-T.}\ \bibnamefont {Yang}}, \bibinfo
  {author} {\bibfnamefont {A.}~\bibnamefont {Laref}}, \ and\ \bibinfo {author}
  {\bibfnamefont {Y.-C.}\ \bibnamefont {Xiong}},\ }\href {\doibase
  10.1103/PhysRevB.105.075430} {\bibfield  {journal} {\bibinfo  {journal}
  {Phys. Rev. B}\ }\textbf {\bibinfo {volume} {105}},\ \bibinfo {pages}
  {075430} (\bibinfo {year} {2022})}\BibitemShut {NoStop}%
\bibitem [{\citenamefont {Takakura}\ \emph {et~al.}(2014)\citenamefont
  {Takakura}, \citenamefont {Noiri}, \citenamefont {Obata}, \citenamefont
  {Otsuka}, \citenamefont {Yoneda}, \citenamefont {Yoshida},\ and\
  \citenamefont {Tarucha}}]{10.1063/1.4869108}%
  \BibitemOpen
  \bibfield  {author} {\bibinfo {author} {\bibfnamefont {T.}~\bibnamefont
  {Takakura}}, \bibinfo {author} {\bibfnamefont {A.}~\bibnamefont {Noiri}},
  \bibinfo {author} {\bibfnamefont {T.}~\bibnamefont {Obata}}, \bibinfo
  {author} {\bibfnamefont {T.}~\bibnamefont {Otsuka}}, \bibinfo {author}
  {\bibfnamefont {J.}~\bibnamefont {Yoneda}}, \bibinfo {author} {\bibfnamefont
  {K.}~\bibnamefont {Yoshida}}, \ and\ \bibinfo {author} {\bibfnamefont
  {S.}~\bibnamefont {Tarucha}},\ }\href {\doibase 10.1063/1.4869108} {\bibfield
   {journal} {\bibinfo  {journal} {Appl. Phys. Lett.}\ }\textbf {\bibinfo
  {volume} {104}},\ \bibinfo {pages} {113109} (\bibinfo {year}
  {2014})}\BibitemShut {NoStop}%
\bibitem [{\citenamefont {Srinivasa}\ \emph {et~al.}(2015)\citenamefont
  {Srinivasa}, \citenamefont {Xu},\ and\ \citenamefont
  {Taylor}}]{PhysRevLett.114.226803}%
  \BibitemOpen
  \bibfield  {author} {\bibinfo {author} {\bibfnamefont {V.}~\bibnamefont
  {Srinivasa}}, \bibinfo {author} {\bibfnamefont {H.}~\bibnamefont {Xu}}, \
  and\ \bibinfo {author} {\bibfnamefont {J.~M.}\ \bibnamefont {Taylor}},\
  }\href {\doibase 10.1103/PhysRevLett.114.226803} {\bibfield  {journal}
  {\bibinfo  {journal} {Phys. Rev. Lett.}\ }\textbf {\bibinfo {volume} {114}},\
  \bibinfo {pages} {226803} (\bibinfo {year} {2015})}\BibitemShut {NoStop}%
\bibitem [{\citenamefont {Volk}\ \emph {et~al.}(2019)\citenamefont {Volk},
  \citenamefont {Chatterjee}, \citenamefont {Ansaloni}, \citenamefont
  {Marcus},\ and\ \citenamefont {Kuemmeth}}]{10.1021/acs.nanolett.9b02149}%
  \BibitemOpen
  \bibfield  {author} {\bibinfo {author} {\bibfnamefont {C.}~\bibnamefont
  {Volk}}, \bibinfo {author} {\bibfnamefont {A.}~\bibnamefont {Chatterjee}},
  \bibinfo {author} {\bibfnamefont {F.}~\bibnamefont {Ansaloni}}, \bibinfo
  {author} {\bibfnamefont {C.~M.}\ \bibnamefont {Marcus}}, \ and\ \bibinfo
  {author} {\bibfnamefont {F.}~\bibnamefont {Kuemmeth}},\ }\href
  {http://dx.doi.org/10.1021/acs.nanolett.9b02149} {\bibfield  {journal}
  {\bibinfo  {journal} {Nano Lett.}\ }\textbf {\bibinfo {volume} {19}},\
  \bibinfo {pages} {5628} (\bibinfo {year} {2019})}\BibitemShut {NoStop}%
\bibitem [{\citenamefont {Leon}\ \emph {et~al.}(2021)\citenamefont {Leon},
  \citenamefont {Yang}, \citenamefont {Hwang}, \citenamefont {Camirand~Lemyre},
  \citenamefont {Tanttu}, \citenamefont {Huang}, \citenamefont {Huang},
  \citenamefont {Hudson}, \citenamefont {Itoh}, \citenamefont {Laucht},
  \citenamefont {Pioro-Ladrière}, \citenamefont {Saraiva},\ and\ \citenamefont
  {Dzurak}}]{Leon.21}%
  \BibitemOpen
  \bibfield  {author} {\bibinfo {author} {\bibfnamefont {R.~C.~C.}\
  \bibnamefont {Leon}}, \bibinfo {author} {\bibfnamefont {C.~H.}\ \bibnamefont
  {Yang}}, \bibinfo {author} {\bibfnamefont {J.~C.~C.}\ \bibnamefont {Hwang}},
  \bibinfo {author} {\bibfnamefont {J.}~\bibnamefont {Camirand~Lemyre}},
  \bibinfo {author} {\bibfnamefont {T.}~\bibnamefont {Tanttu}}, \bibinfo
  {author} {\bibfnamefont {W.}~\bibnamefont {Huang}}, \bibinfo {author}
  {\bibfnamefont {J.~Y.}\ \bibnamefont {Huang}}, \bibinfo {author}
  {\bibfnamefont {F.~E.}\ \bibnamefont {Hudson}}, \bibinfo {author}
  {\bibfnamefont {K.~M.}\ \bibnamefont {Itoh}}, \bibinfo {author}
  {\bibfnamefont {A.}~\bibnamefont {Laucht}}, \bibinfo {author} {\bibfnamefont
  {M.}~\bibnamefont {Pioro-Ladrière}}, \bibinfo {author} {\bibfnamefont
  {A.}~\bibnamefont {Saraiva}}, \ and\ \bibinfo {author} {\bibfnamefont
  {A.~S.}\ \bibnamefont {Dzurak}},\ }\href
  {http://dx.doi.org/10.1038/s41467-021-23437-w} {\bibfield  {journal}
  {\bibinfo  {journal} {Nat. Commun.}\ }\textbf {\bibinfo {volume} {12}},\
  \bibinfo {pages} {3228} (\bibinfo {year} {2021})}\BibitemShut {NoStop}%
\bibitem [{\citenamefont {Potts}\ \emph {et~al.}(2021)\citenamefont {Potts},
  \citenamefont {Josefi}, \citenamefont {Chen}, \citenamefont {Lehmann},
  \citenamefont {Dick}, \citenamefont {Leijnse}, \citenamefont {Reimann},
  \citenamefont {Bengtsson},\ and\ \citenamefont
  {Thelander}}]{PhysRevB.104.L081409}%
  \BibitemOpen
  \bibfield  {author} {\bibinfo {author} {\bibfnamefont {H.}~\bibnamefont
  {Potts}}, \bibinfo {author} {\bibfnamefont {J.}~\bibnamefont {Josefi}},
  \bibinfo {author} {\bibfnamefont {I.-J.}\ \bibnamefont {Chen}}, \bibinfo
  {author} {\bibfnamefont {S.}~\bibnamefont {Lehmann}}, \bibinfo {author}
  {\bibfnamefont {K.~A.}\ \bibnamefont {Dick}}, \bibinfo {author}
  {\bibfnamefont {M.}~\bibnamefont {Leijnse}}, \bibinfo {author} {\bibfnamefont
  {S.~M.}\ \bibnamefont {Reimann}}, \bibinfo {author} {\bibfnamefont
  {J.}~\bibnamefont {Bengtsson}}, \ and\ \bibinfo {author} {\bibfnamefont
  {C.}~\bibnamefont {Thelander}},\ }\href
  {https://link.aps.org/doi/10.1103/PhysRevB.104.L081409} {\bibfield  {journal}
  {\bibinfo  {journal} {Phys. Rev. B}\ }\textbf {\bibinfo {volume} {104}},\
  \bibinfo {pages} {L081409} (\bibinfo {year} {2021})}\BibitemShut {NoStop}%
\bibitem [{\citenamefont {Mehl}\ and\ \citenamefont
  {DiVincenzo}(2014)}]{PhysRevB.90.195424}%
  \BibitemOpen
  \bibfield  {author} {\bibinfo {author} {\bibfnamefont {S.}~\bibnamefont
  {Mehl}}\ and\ \bibinfo {author} {\bibfnamefont {D.~P.}\ \bibnamefont
  {DiVincenzo}},\ }\href {https://link.aps.org/doi/10.1103/PhysRevB.90.195424}
  {\bibfield  {journal} {\bibinfo  {journal} {Phys. Rev. B}\ }\textbf {\bibinfo
  {volume} {90}},\ \bibinfo {pages} {195424} (\bibinfo {year}
  {2014})}\BibitemShut {NoStop}%
\bibitem [{\citenamefont {Kiyama}\ \emph {et~al.}(2021)\citenamefont {Kiyama},
  \citenamefont {Yoshimi}, \citenamefont {Kato}, \citenamefont {Nakajima},
  \citenamefont {Oiwa},\ and\ \citenamefont
  {Tarucha}}]{PhysRevLett.127.086802}%
  \BibitemOpen
  \bibfield  {author} {\bibinfo {author} {\bibfnamefont {H.}~\bibnamefont
  {Kiyama}}, \bibinfo {author} {\bibfnamefont {K.}~\bibnamefont {Yoshimi}},
  \bibinfo {author} {\bibfnamefont {T.}~\bibnamefont {Kato}}, \bibinfo {author}
  {\bibfnamefont {T.}~\bibnamefont {Nakajima}}, \bibinfo {author}
  {\bibfnamefont {A.}~\bibnamefont {Oiwa}}, \ and\ \bibinfo {author}
  {\bibfnamefont {S.}~\bibnamefont {Tarucha}},\ }\href
  {https://link.aps.org/doi/10.1103/PhysRevLett.127.086802} {\bibfield
  {journal} {\bibinfo  {journal} {Phys. Rev. Lett.}\ }\textbf {\bibinfo
  {volume} {127}},\ \bibinfo {pages} {086802} (\bibinfo {year}
  {2021})}\BibitemShut {NoStop}%
\bibitem [{\citenamefont {Malinowski}\ \emph {et~al.}(2018)\citenamefont
  {Malinowski}, \citenamefont {Martins}, \citenamefont {Smith}, \citenamefont
  {Bartlett}, \citenamefont {Doherty}, \citenamefont {Nissen}, \citenamefont
  {Fallahi}, \citenamefont {Gardner}, \citenamefont {Manfra}, \citenamefont
  {Marcus},\ and\ \citenamefont {Kuemmeth}}]{PhysRevX.8.011045}%
  \BibitemOpen
  \bibfield  {author} {\bibinfo {author} {\bibfnamefont {F.~K.}\ \bibnamefont
  {Malinowski}}, \bibinfo {author} {\bibfnamefont {F.}~\bibnamefont {Martins}},
  \bibinfo {author} {\bibfnamefont {T.~B.}\ \bibnamefont {Smith}}, \bibinfo
  {author} {\bibfnamefont {S.~D.}\ \bibnamefont {Bartlett}}, \bibinfo {author}
  {\bibfnamefont {A.~C.}\ \bibnamefont {Doherty}}, \bibinfo {author}
  {\bibfnamefont {P.~D.}\ \bibnamefont {Nissen}}, \bibinfo {author}
  {\bibfnamefont {S.}~\bibnamefont {Fallahi}}, \bibinfo {author} {\bibfnamefont
  {G.~C.}\ \bibnamefont {Gardner}}, \bibinfo {author} {\bibfnamefont {M.~J.}\
  \bibnamefont {Manfra}}, \bibinfo {author} {\bibfnamefont {C.~M.}\
  \bibnamefont {Marcus}}, \ and\ \bibinfo {author} {\bibfnamefont
  {F.}~\bibnamefont {Kuemmeth}},\ }\href
  {https://link.aps.org/doi/10.1103/PhysRevX.8.011045} {\bibfield  {journal}
  {\bibinfo  {journal} {Phys. Rev. X}\ }\textbf {\bibinfo {volume} {8}},\
  \bibinfo {pages} {011045} (\bibinfo {year} {2018})}\BibitemShut {NoStop}%
\bibitem [{\citenamefont {Martins}\ \emph {et~al.}(2017)\citenamefont
  {Martins}, \citenamefont {Malinowski}, \citenamefont {Nissen}, \citenamefont
  {Fallahi}, \citenamefont {Gardner}, \citenamefont {Manfra}, \citenamefont
  {Marcus},\ and\ \citenamefont {Kuemmeth}}]{PhysRevLett.119.227701}%
  \BibitemOpen
  \bibfield  {author} {\bibinfo {author} {\bibfnamefont {F.}~\bibnamefont
  {Martins}}, \bibinfo {author} {\bibfnamefont {F.~K.}\ \bibnamefont
  {Malinowski}}, \bibinfo {author} {\bibfnamefont {P.~D.}\ \bibnamefont
  {Nissen}}, \bibinfo {author} {\bibfnamefont {S.}~\bibnamefont {Fallahi}},
  \bibinfo {author} {\bibfnamefont {G.~C.}\ \bibnamefont {Gardner}}, \bibinfo
  {author} {\bibfnamefont {M.~J.}\ \bibnamefont {Manfra}}, \bibinfo {author}
  {\bibfnamefont {C.~M.}\ \bibnamefont {Marcus}}, \ and\ \bibinfo {author}
  {\bibfnamefont {F.}~\bibnamefont {Kuemmeth}},\ }\href
  {https://link.aps.org/doi/10.1103/PhysRevLett.119.227701} {\bibfield
  {journal} {\bibinfo  {journal} {Phys. Rev. Lett.}\ }\textbf {\bibinfo
  {volume} {119}},\ \bibinfo {pages} {227701} (\bibinfo {year}
  {2017})}\BibitemShut {NoStop}%
\bibitem [{\citenamefont {Kouwenhoven}\ \emph {et~al.}(1997)\citenamefont
  {Kouwenhoven}, \citenamefont {Oosterkamp}, \citenamefont {Danoesastro},
  \citenamefont {Eto}, \citenamefont {Austing}, \citenamefont {Honda},\ and\
  \citenamefont {Tarucha}}]{science.278.5344.1788}%
  \BibitemOpen
  \bibfield  {author} {\bibinfo {author} {\bibfnamefont {L.~P.}\ \bibnamefont
  {Kouwenhoven}}, \bibinfo {author} {\bibfnamefont {T.}~\bibnamefont
  {Oosterkamp}}, \bibinfo {author} {\bibfnamefont {M.}~\bibnamefont
  {Danoesastro}}, \bibinfo {author} {\bibfnamefont {M.}~\bibnamefont {Eto}},
  \bibinfo {author} {\bibfnamefont {D.}~\bibnamefont {Austing}}, \bibinfo
  {author} {\bibfnamefont {T.}~\bibnamefont {Honda}}, \ and\ \bibinfo {author}
  {\bibfnamefont {S.}~\bibnamefont {Tarucha}},\ }\href
  {http://science.sciencemag.org/content/278/5344/1788} {\bibfield  {journal}
  {\bibinfo  {journal} {Science}\ }\textbf {\bibinfo {volume} {278}},\ \bibinfo
  {pages} {1788} (\bibinfo {year} {1997})}\BibitemShut {NoStop}%
\bibitem [{\citenamefont {Higginbotham}\ \emph {et~al.}(2014)\citenamefont
  {Higginbotham}, \citenamefont {Kuemmeth}, \citenamefont {Hanson},
  \citenamefont {Gossard},\ and\ \citenamefont
  {Marcus}}]{PhysRevLett.112.026801}%
  \BibitemOpen
  \bibfield  {author} {\bibinfo {author} {\bibfnamefont {A.~P.}\ \bibnamefont
  {Higginbotham}}, \bibinfo {author} {\bibfnamefont {F.}~\bibnamefont
  {Kuemmeth}}, \bibinfo {author} {\bibfnamefont {M.~P.}\ \bibnamefont
  {Hanson}}, \bibinfo {author} {\bibfnamefont {A.~C.}\ \bibnamefont {Gossard}},
  \ and\ \bibinfo {author} {\bibfnamefont {C.~M.}\ \bibnamefont {Marcus}},\
  }\href {https://link.aps.org/doi/10.1103/PhysRevLett.112.026801} {\bibfield
  {journal} {\bibinfo  {journal} {Phys. Rev. Lett.}\ }\textbf {\bibinfo
  {volume} {112}},\ \bibinfo {pages} {026801} (\bibinfo {year}
  {2014})}\BibitemShut {NoStop}%
\bibitem [{\citenamefont {Bakker}\ \emph {et~al.}(2015)\citenamefont {Bakker},
  \citenamefont {Mehl}, \citenamefont {Hiltunen}, \citenamefont {Harju},\ and\
  \citenamefont {DiVincenzo}}]{PhysRevB.91.155425}%
  \BibitemOpen
  \bibfield  {author} {\bibinfo {author} {\bibfnamefont {M.~A.}\ \bibnamefont
  {Bakker}}, \bibinfo {author} {\bibfnamefont {S.}~\bibnamefont {Mehl}},
  \bibinfo {author} {\bibfnamefont {T.}~\bibnamefont {Hiltunen}}, \bibinfo
  {author} {\bibfnamefont {A.}~\bibnamefont {Harju}}, \ and\ \bibinfo {author}
  {\bibfnamefont {D.~P.}\ \bibnamefont {DiVincenzo}},\ }\href
  {https://link.aps.org/doi/10.1103/PhysRevB.91.155425} {\bibfield  {journal}
  {\bibinfo  {journal} {Phys. Rev. B}\ }\textbf {\bibinfo {volume} {91}},\
  \bibinfo {pages} {155425} (\bibinfo {year} {2015})}\BibitemShut {NoStop}%
\bibitem [{\citenamefont {Mehl}\ and\ \citenamefont
  {DiVincenzo}(2013)}]{PhysRevB.88.161408}%
  \BibitemOpen
  \bibfield  {author} {\bibinfo {author} {\bibfnamefont {S.}~\bibnamefont
  {Mehl}}\ and\ \bibinfo {author} {\bibfnamefont {D.~P.}\ \bibnamefont
  {DiVincenzo}},\ }\href {\doibase 10.1103/PhysRevB.88.161408} {\bibfield
  {journal} {\bibinfo  {journal} {Phys. Rev. B}\ }\textbf {\bibinfo {volume}
  {88}},\ \bibinfo {pages} {161408} (\bibinfo {year} {2013})}\BibitemShut
  {NoStop}%
\bibitem [{\citenamefont {Lundberg}\ \emph {et~al.}(2020)\citenamefont
  {Lundberg}, \citenamefont {Li}, \citenamefont {Hutin}, \citenamefont
  {Bertrand}, \citenamefont {Ibberson}, \citenamefont {Lee}, \citenamefont
  {Niegemann}, \citenamefont {Urdampilleta}, \citenamefont {Stelmashenko},
  \citenamefont {Meunier}, \citenamefont {Robinson}, \citenamefont {Ibberson},
  \citenamefont {Vinet}, \citenamefont {Niquet},\ and\ \citenamefont
  {Gonzalez-Zalba}}]{PhysRevX.10.041010}%
  \BibitemOpen
  \bibfield  {author} {\bibinfo {author} {\bibfnamefont {T.}~\bibnamefont
  {Lundberg}}, \bibinfo {author} {\bibfnamefont {J.}~\bibnamefont {Li}},
  \bibinfo {author} {\bibfnamefont {L.}~\bibnamefont {Hutin}}, \bibinfo
  {author} {\bibfnamefont {B.}~\bibnamefont {Bertrand}}, \bibinfo {author}
  {\bibfnamefont {D.~J.}\ \bibnamefont {Ibberson}}, \bibinfo {author}
  {\bibfnamefont {C.-M.}\ \bibnamefont {Lee}}, \bibinfo {author} {\bibfnamefont
  {D.~J.}\ \bibnamefont {Niegemann}}, \bibinfo {author} {\bibfnamefont
  {M.}~\bibnamefont {Urdampilleta}}, \bibinfo {author} {\bibfnamefont
  {N.}~\bibnamefont {Stelmashenko}}, \bibinfo {author} {\bibfnamefont
  {T.}~\bibnamefont {Meunier}}, \bibinfo {author} {\bibfnamefont {J.~W.~A.}\
  \bibnamefont {Robinson}}, \bibinfo {author} {\bibfnamefont {L.}~\bibnamefont
  {Ibberson}}, \bibinfo {author} {\bibfnamefont {M.}~\bibnamefont {Vinet}},
  \bibinfo {author} {\bibfnamefont {Y.-M.}\ \bibnamefont {Niquet}}, \ and\
  \bibinfo {author} {\bibfnamefont {M.~F.}\ \bibnamefont {Gonzalez-Zalba}},\
  }\href {https://link.aps.org/doi/10.1103/PhysRevX.10.041010} {\bibfield
  {journal} {\bibinfo  {journal} {Phys. Rev. X}\ }\textbf {\bibinfo {volume}
  {10}},\ \bibinfo {pages} {041010} (\bibinfo {year} {2020})}\BibitemShut
  {NoStop}%
\bibitem [{\citenamefont {Deng}\ \emph {et~al.}(2018)\citenamefont {Deng},
  \citenamefont {Calderon-Vargas}, \citenamefont {Mayhall},\ and\ \citenamefont
  {Barnes}}]{PhysRevB.97.245301}%
  \BibitemOpen
  \bibfield  {author} {\bibinfo {author} {\bibfnamefont {K.}~\bibnamefont
  {Deng}}, \bibinfo {author} {\bibfnamefont {F.~A.}\ \bibnamefont
  {Calderon-Vargas}}, \bibinfo {author} {\bibfnamefont {N.~J.}\ \bibnamefont
  {Mayhall}}, \ and\ \bibinfo {author} {\bibfnamefont {E.}~\bibnamefont
  {Barnes}},\ }\href {https://link.aps.org/doi/10.1103/PhysRevB.97.245301}
  {\bibfield  {journal} {\bibinfo  {journal} {Phys. Rev. B}\ }\textbf {\bibinfo
  {volume} {97}},\ \bibinfo {pages} {245301} (\bibinfo {year}
  {2018})}\BibitemShut {NoStop}%
\bibitem [{\citenamefont {Chan}\ and\ \citenamefont
  {Wang}(2022{\natexlab{a}})}]{chan2022microscopic}%
  \BibitemOpen
  \bibfield  {author} {\bibinfo {author} {\bibfnamefont {G.~X.}\ \bibnamefont
  {Chan}}\ and\ \bibinfo {author} {\bibfnamefont {X.}~\bibnamefont {Wang}},\
  }\href {https://arxiv.org/abs/2202.02308} {\bibfield  {journal} {\bibinfo
  {journal} {arXiv preprint arXiv:2202.02308}\ } (\bibinfo {year}
  {2022}{\natexlab{a}})}\BibitemShut {NoStop}%
\bibitem [{\citenamefont {Chan}\ and\ \citenamefont
  {Wang}(2022{\natexlab{b}})}]{chan2022robust}%
  \BibitemOpen
  \bibfield  {author} {\bibinfo {author} {\bibfnamefont {G.~X.}\ \bibnamefont
  {Chan}}\ and\ \bibinfo {author} {\bibfnamefont {X.}~\bibnamefont {Wang}},\
  }\href {https://arxiv.org/abs/2201.01583} {\bibfield  {journal} {\bibinfo
  {journal} {arXiv preprint arXiv:2201.01583}\ } (\bibinfo {year}
  {2022}{\natexlab{b}})}\BibitemShut {NoStop}%
\bibitem [{\citenamefont {Zhao}\ and\ \citenamefont
  {Hu}(2018)}]{zhao2018toward}%
  \BibitemOpen
  \bibfield  {author} {\bibinfo {author} {\bibfnamefont {X.}~\bibnamefont
  {Zhao}}\ and\ \bibinfo {author} {\bibfnamefont {X.}~\bibnamefont {Hu}},\
  }\href {https://doi.org/10.1038/s41598-018-31879-4} {\bibfield  {journal}
  {\bibinfo  {journal} {Sci. Rep.}\ }\textbf {\bibinfo {volume} {8}},\ \bibinfo
  {pages} {13968} (\bibinfo {year} {2018})}\BibitemShut {NoStop}%
\bibitem [{\citenamefont {Kornich}\ \emph {et~al.}(2018)\citenamefont
  {Kornich}, \citenamefont {Kloeffel},\ and\ \citenamefont
  {Loss}}]{kornich2018phonon}%
  \BibitemOpen
  \bibfield  {author} {\bibinfo {author} {\bibfnamefont {V.}~\bibnamefont
  {Kornich}}, \bibinfo {author} {\bibfnamefont {C.}~\bibnamefont {Kloeffel}}, \
  and\ \bibinfo {author} {\bibfnamefont {D.}~\bibnamefont {Loss}},\ }\href
  {https://doi.org/10.1038/s41598-021-94621-7} {\bibfield  {journal} {\bibinfo
  {journal} {Quantum}\ }\textbf {\bibinfo {volume} {2}},\ \bibinfo {pages} {70}
  (\bibinfo {year} {2018})}\BibitemShut {NoStop}%
\bibitem [{\citenamefont {Kornich}\ \emph {et~al.}(2014)\citenamefont
  {Kornich}, \citenamefont {Kloeffel},\ and\ \citenamefont
  {Loss}}]{PhysRevB.89.085410}%
  \BibitemOpen
  \bibfield  {author} {\bibinfo {author} {\bibfnamefont {V.}~\bibnamefont
  {Kornich}}, \bibinfo {author} {\bibfnamefont {C.}~\bibnamefont {Kloeffel}}, \
  and\ \bibinfo {author} {\bibfnamefont {D.}~\bibnamefont {Loss}},\ }\href
  {\doibase 10.1103/PhysRevB.89.085410} {\bibfield  {journal} {\bibinfo
  {journal} {Phys. Rev. B}\ }\textbf {\bibinfo {volume} {89}},\ \bibinfo
  {pages} {085410} (\bibinfo {year} {2014})}\BibitemShut {NoStop}%
\bibitem [{\citenamefont {Hu}(2011)}]{PhysRevB.83.165322}%
  \BibitemOpen
  \bibfield  {author} {\bibinfo {author} {\bibfnamefont {X.}~\bibnamefont
  {Hu}},\ }\href {https://link.aps.org/doi/10.1103/PhysRevB.83.165322}
  {\bibfield  {journal} {\bibinfo  {journal} {Phys. Rev. B}\ }\textbf {\bibinfo
  {volume} {83}},\ \bibinfo {pages} {165322} (\bibinfo {year}
  {2011})}\BibitemShut {NoStop}%
\bibitem [{\citenamefont {Mozyrsky}\ \emph {et~al.}(2002)\citenamefont
  {Mozyrsky}, \citenamefont {Kogan}, \citenamefont {Gorshkov},\ and\
  \citenamefont {Berman}}]{PhysRevB.65.245213}%
  \BibitemOpen
  \bibfield  {author} {\bibinfo {author} {\bibfnamefont {D.}~\bibnamefont
  {Mozyrsky}}, \bibinfo {author} {\bibfnamefont {S.}~\bibnamefont {Kogan}},
  \bibinfo {author} {\bibfnamefont {V.~N.}\ \bibnamefont {Gorshkov}}, \ and\
  \bibinfo {author} {\bibfnamefont {G.~P.}\ \bibnamefont {Berman}},\ }\href
  {https://link.aps.org/doi/10.1103/PhysRevB.65.245213} {\bibfield  {journal}
  {\bibinfo  {journal} {Phys. Rev. B}\ }\textbf {\bibinfo {volume} {65}},\
  \bibinfo {pages} {245213} (\bibinfo {year} {2002})}\BibitemShut {NoStop}%
\bibitem [{\citenamefont {Dial}\ \emph {et~al.}(2013)\citenamefont {Dial},
  \citenamefont {Shulman}, \citenamefont {Harvey}, \citenamefont {Bluhm},
  \citenamefont {Umansky},\ and\ \citenamefont
  {Yacoby}}]{PhysRevLett.110.146804}%
  \BibitemOpen
  \bibfield  {author} {\bibinfo {author} {\bibfnamefont {O.~E.}\ \bibnamefont
  {Dial}}, \bibinfo {author} {\bibfnamefont {M.~D.}\ \bibnamefont {Shulman}},
  \bibinfo {author} {\bibfnamefont {S.~P.}\ \bibnamefont {Harvey}}, \bibinfo
  {author} {\bibfnamefont {H.}~\bibnamefont {Bluhm}}, \bibinfo {author}
  {\bibfnamefont {V.}~\bibnamefont {Umansky}}, \ and\ \bibinfo {author}
  {\bibfnamefont {A.}~\bibnamefont {Yacoby}},\ }\href
  {https://link.aps.org/doi/10.1103/PhysRevLett.110.146804} {\bibfield
  {journal} {\bibinfo  {journal} {Phys. Rev. Lett.}\ }\textbf {\bibinfo
  {volume} {110}},\ \bibinfo {pages} {146804} (\bibinfo {year}
  {2013})}\BibitemShut {NoStop}%
\bibitem [{\citenamefont {Hu}\ and\ \citenamefont
  {Das~Sarma}(2006)}]{PhysRevLett.96.100501}%
  \BibitemOpen
  \bibfield  {author} {\bibinfo {author} {\bibfnamefont {X.}~\bibnamefont
  {Hu}}\ and\ \bibinfo {author} {\bibfnamefont {S.}~\bibnamefont {Das~Sarma}},\
  }\href {https://link.aps.org/doi/10.1103/PhysRevLett.96.100501} {\bibfield
  {journal} {\bibinfo  {journal} {Phys. Rev. Lett.}\ }\textbf {\bibinfo
  {volume} {96}},\ \bibinfo {pages} {100501} (\bibinfo {year}
  {2006})}\BibitemShut {NoStop}%
\bibitem [{\citenamefont {Yoneda}\ \emph {et~al.}(2018)\citenamefont {Yoneda},
  \citenamefont {Takeda}, \citenamefont {Otsuka}, \citenamefont {Nakajima},
  \citenamefont {Delbecq}, \citenamefont {Allison}, \citenamefont {Honda},
  \citenamefont {Kodera}, \citenamefont {Oda}, \citenamefont {Hoshi} \emph
  {et~al.}}]{yoneda2018quantum}%
  \BibitemOpen
  \bibfield  {author} {\bibinfo {author} {\bibfnamefont {J.}~\bibnamefont
  {Yoneda}}, \bibinfo {author} {\bibfnamefont {K.}~\bibnamefont {Takeda}},
  \bibinfo {author} {\bibfnamefont {T.}~\bibnamefont {Otsuka}}, \bibinfo
  {author} {\bibfnamefont {T.}~\bibnamefont {Nakajima}}, \bibinfo {author}
  {\bibfnamefont {M.~R.}\ \bibnamefont {Delbecq}}, \bibinfo {author}
  {\bibfnamefont {G.}~\bibnamefont {Allison}}, \bibinfo {author} {\bibfnamefont
  {T.}~\bibnamefont {Honda}}, \bibinfo {author} {\bibfnamefont
  {T.}~\bibnamefont {Kodera}}, \bibinfo {author} {\bibfnamefont
  {S.}~\bibnamefont {Oda}}, \bibinfo {author} {\bibfnamefont {Y.}~\bibnamefont
  {Hoshi}},  \emph {et~al.},\ }\href
  {https://doi.org/10.1038/s41565-017-0014-x} {\bibfield  {journal} {\bibinfo
  {journal} {Nat. Nanotechnol.}\ }\textbf {\bibinfo {volume} {13}},\ \bibinfo
  {pages} {102} (\bibinfo {year} {2018})}\BibitemShut {NoStop}%
\bibitem [{\citenamefont {Shulman}\ \emph {et~al.}(2012)\citenamefont
  {Shulman}, \citenamefont {Dial}, \citenamefont {Harvey}, \citenamefont
  {Bluhm}, \citenamefont {Umansky},\ and\ \citenamefont
  {Yacoby}}]{doi:10.1126/science.1217692}%
  \BibitemOpen
  \bibfield  {author} {\bibinfo {author} {\bibfnamefont {M.~D.}\ \bibnamefont
  {Shulman}}, \bibinfo {author} {\bibfnamefont {O.~E.}\ \bibnamefont {Dial}},
  \bibinfo {author} {\bibfnamefont {S.~P.}\ \bibnamefont {Harvey}}, \bibinfo
  {author} {\bibfnamefont {H.}~\bibnamefont {Bluhm}}, \bibinfo {author}
  {\bibfnamefont {V.}~\bibnamefont {Umansky}}, \ and\ \bibinfo {author}
  {\bibfnamefont {A.}~\bibnamefont {Yacoby}},\ }\href
  {https://www.science.org/doi/abs/10.1126/science.1217692} {\bibfield
  {journal} {\bibinfo  {journal} {Science}\ }\textbf {\bibinfo {volume}
  {336}},\ \bibinfo {pages} {202} (\bibinfo {year} {2012})}\BibitemShut
  {NoStop}%
\bibitem [{\citenamefont {Reed}\ \emph {et~al.}(2016)\citenamefont {Reed},
  \citenamefont {Maune}, \citenamefont {Andrews}, \citenamefont {Borselli},
  \citenamefont {Eng}, \citenamefont {Jura}, \citenamefont {Kiselev},
  \citenamefont {Ladd}, \citenamefont {Merkel}, \citenamefont {Milosavljevic},
  \citenamefont {Pritchett}, \citenamefont {Rakher}, \citenamefont {Ross},
  \citenamefont {Schmitz}, \citenamefont {Smith}, \citenamefont {Wright},
  \citenamefont {Gyure},\ and\ \citenamefont
  {Hunter}}]{PhysRevLett.116.110402}%
  \BibitemOpen
  \bibfield  {author} {\bibinfo {author} {\bibfnamefont {M.~D.}\ \bibnamefont
  {Reed}}, \bibinfo {author} {\bibfnamefont {B.~M.}\ \bibnamefont {Maune}},
  \bibinfo {author} {\bibfnamefont {R.~W.}\ \bibnamefont {Andrews}}, \bibinfo
  {author} {\bibfnamefont {M.~G.}\ \bibnamefont {Borselli}}, \bibinfo {author}
  {\bibfnamefont {K.}~\bibnamefont {Eng}}, \bibinfo {author} {\bibfnamefont
  {M.~P.}\ \bibnamefont {Jura}}, \bibinfo {author} {\bibfnamefont {A.~A.}\
  \bibnamefont {Kiselev}}, \bibinfo {author} {\bibfnamefont {T.~D.}\
  \bibnamefont {Ladd}}, \bibinfo {author} {\bibfnamefont {S.~T.}\ \bibnamefont
  {Merkel}}, \bibinfo {author} {\bibfnamefont {I.}~\bibnamefont
  {Milosavljevic}}, \bibinfo {author} {\bibfnamefont {E.~J.}\ \bibnamefont
  {Pritchett}}, \bibinfo {author} {\bibfnamefont {M.~T.}\ \bibnamefont
  {Rakher}}, \bibinfo {author} {\bibfnamefont {R.~S.}\ \bibnamefont {Ross}},
  \bibinfo {author} {\bibfnamefont {A.~E.}\ \bibnamefont {Schmitz}}, \bibinfo
  {author} {\bibfnamefont {A.}~\bibnamefont {Smith}}, \bibinfo {author}
  {\bibfnamefont {J.~A.}\ \bibnamefont {Wright}}, \bibinfo {author}
  {\bibfnamefont {M.~F.}\ \bibnamefont {Gyure}}, \ and\ \bibinfo {author}
  {\bibfnamefont {A.~T.}\ \bibnamefont {Hunter}},\ }\href
  {https://link.aps.org/doi/10.1103/PhysRevLett.116.110402} {\bibfield
  {journal} {\bibinfo  {journal} {Phys. Rev. Lett.}\ }\textbf {\bibinfo
  {volume} {116}},\ \bibinfo {pages} {110402} (\bibinfo {year}
  {2016})}\BibitemShut {NoStop}%
\bibitem [{\citenamefont {Huang}(2021)}]{huang2021dephasing}%
  \BibitemOpen
  \bibfield  {author} {\bibinfo {author} {\bibfnamefont {P.}~\bibnamefont
  {Huang}},\ }\href {https://arxiv.org/abs/2109.02261} {\bibfield  {journal}
  {\bibinfo  {journal} {Adv. Quantum Technol.}\ }\textbf {\bibinfo {volume}
  {4}},\ \bibinfo {pages} {2100018} (\bibinfo {year} {2021})}\BibitemShut
  {NoStop}%
\bibitem [{\citenamefont {Taylor}\ \emph {et~al.}(2007)\citenamefont {Taylor},
  \citenamefont {Petta}, \citenamefont {Johnson}, \citenamefont {Yacoby},
  \citenamefont {Marcus},\ and\ \citenamefont {Lukin}}]{PhysRevB.76.035315}%
  \BibitemOpen
  \bibfield  {author} {\bibinfo {author} {\bibfnamefont {J.~M.}\ \bibnamefont
  {Taylor}}, \bibinfo {author} {\bibfnamefont {J.~R.}\ \bibnamefont {Petta}},
  \bibinfo {author} {\bibfnamefont {A.~C.}\ \bibnamefont {Johnson}}, \bibinfo
  {author} {\bibfnamefont {A.}~\bibnamefont {Yacoby}}, \bibinfo {author}
  {\bibfnamefont {C.~M.}\ \bibnamefont {Marcus}}, \ and\ \bibinfo {author}
  {\bibfnamefont {M.~D.}\ \bibnamefont {Lukin}},\ }\href
  {https://link.aps.org/doi/10.1103/PhysRevB.76.035315} {\bibfield  {journal}
  {\bibinfo  {journal} {Phys. Rev. B}\ }\textbf {\bibinfo {volume} {76}},\
  \bibinfo {pages} {035315} (\bibinfo {year} {2007})}\BibitemShut {NoStop}%
\bibitem [{\citenamefont {Gamble}\ \emph {et~al.}(2012)\citenamefont {Gamble},
  \citenamefont {Friesen}, \citenamefont {Coppersmith},\ and\ \citenamefont
  {Hu}}]{PhysRevB.86.035302}%
  \BibitemOpen
  \bibfield  {author} {\bibinfo {author} {\bibfnamefont {J.~K.}\ \bibnamefont
  {Gamble}}, \bibinfo {author} {\bibfnamefont {M.}~\bibnamefont {Friesen}},
  \bibinfo {author} {\bibfnamefont {S.~N.}\ \bibnamefont {Coppersmith}}, \ and\
  \bibinfo {author} {\bibfnamefont {X.}~\bibnamefont {Hu}},\ }\href
  {https://link.aps.org/doi/10.1103/PhysRevB.86.035302} {\bibfield  {journal}
  {\bibinfo  {journal} {Phys. Rev. B}\ }\textbf {\bibinfo {volume} {86}},\
  \bibinfo {pages} {035302} (\bibinfo {year} {2012})}\BibitemShut {NoStop}%
\bibitem [{\citenamefont {Madzik}\ \emph {et~al.}(2020)\citenamefont {Madzik},
  \citenamefont {Ladd}, \citenamefont {Hudson}, \citenamefont {Itoh},
  \citenamefont {Jakob}, \citenamefont {Johnson}, \citenamefont {McCallum},
  \citenamefont {Jamieson}, \citenamefont {Dzurak}, \citenamefont {Laucht},\
  and\ \citenamefont {Morello}}]{Madzik.20}%
  \BibitemOpen
  \bibfield  {author} {\bibinfo {author} {\bibfnamefont {M.~T.}\ \bibnamefont
  {Madzik}}, \bibinfo {author} {\bibfnamefont {T.~D.}\ \bibnamefont {Ladd}},
  \bibinfo {author} {\bibfnamefont {F.~E.}\ \bibnamefont {Hudson}}, \bibinfo
  {author} {\bibfnamefont {K.~M.}\ \bibnamefont {Itoh}}, \bibinfo {author}
  {\bibfnamefont {A.~M.}\ \bibnamefont {Jakob}}, \bibinfo {author}
  {\bibfnamefont {B.~C.}\ \bibnamefont {Johnson}}, \bibinfo {author}
  {\bibfnamefont {J.~C.}\ \bibnamefont {McCallum}}, \bibinfo {author}
  {\bibfnamefont {D.~N.}\ \bibnamefont {Jamieson}}, \bibinfo {author}
  {\bibfnamefont {A.~S.}\ \bibnamefont {Dzurak}}, \bibinfo {author}
  {\bibfnamefont {A.}~\bibnamefont {Laucht}}, \ and\ \bibinfo {author}
  {\bibfnamefont {A.}~\bibnamefont {Morello}},\ }\href
  {http://dx.doi.org/10.1126/sciadv.aba3442} {\bibfield  {journal} {\bibinfo
  {journal} {Sci. Adv.}\ }\textbf {\bibinfo {volume} {6}} (\bibinfo {year}
  {2020})}\BibitemShut {NoStop}%
\bibitem [{\citenamefont {Benito}\ \emph {et~al.}(2019)\citenamefont {Benito},
  \citenamefont {Croot}, \citenamefont {Adelsberger}, \citenamefont {Putz},
  \citenamefont {Mi}, \citenamefont {Petta},\ and\ \citenamefont
  {Burkard}}]{PhysRevB.100.125430}%
  \BibitemOpen
  \bibfield  {author} {\bibinfo {author} {\bibfnamefont {M.}~\bibnamefont
  {Benito}}, \bibinfo {author} {\bibfnamefont {X.}~\bibnamefont {Croot}},
  \bibinfo {author} {\bibfnamefont {C.}~\bibnamefont {Adelsberger}}, \bibinfo
  {author} {\bibfnamefont {S.}~\bibnamefont {Putz}}, \bibinfo {author}
  {\bibfnamefont {X.}~\bibnamefont {Mi}}, \bibinfo {author} {\bibfnamefont
  {J.~R.}\ \bibnamefont {Petta}}, \ and\ \bibinfo {author} {\bibfnamefont
  {G.}~\bibnamefont {Burkard}},\ }\href {\doibase 10.1103/PhysRevB.100.125430}
  {\bibfield  {journal} {\bibinfo  {journal} {Phys. Rev. B}\ }\textbf {\bibinfo
  {volume} {100}},\ \bibinfo {pages} {125430} (\bibinfo {year}
  {2019})}\BibitemShut {NoStop}%
\bibitem [{\citenamefont {Yoneda}\ \emph {et~al.}(2021)\citenamefont {Yoneda},
  \citenamefont {Huang}, \citenamefont {Feng}, \citenamefont {Yang},
  \citenamefont {Chan}, \citenamefont {Tanttu}, \citenamefont {Gilbert},
  \citenamefont {Leon}, \citenamefont {Hudson}, \citenamefont {Itoh} \emph
  {et~al.}}]{yoneda2021coherent}%
  \BibitemOpen
  \bibfield  {author} {\bibinfo {author} {\bibfnamefont {J.}~\bibnamefont
  {Yoneda}}, \bibinfo {author} {\bibfnamefont {W.}~\bibnamefont {Huang}},
  \bibinfo {author} {\bibfnamefont {M.}~\bibnamefont {Feng}}, \bibinfo {author}
  {\bibfnamefont {C.~H.}\ \bibnamefont {Yang}}, \bibinfo {author}
  {\bibfnamefont {K.~W.}\ \bibnamefont {Chan}}, \bibinfo {author}
  {\bibfnamefont {T.}~\bibnamefont {Tanttu}}, \bibinfo {author} {\bibfnamefont
  {W.}~\bibnamefont {Gilbert}}, \bibinfo {author} {\bibfnamefont
  {R.}~\bibnamefont {Leon}}, \bibinfo {author} {\bibfnamefont {F.}~\bibnamefont
  {Hudson}}, \bibinfo {author} {\bibfnamefont {K.}~\bibnamefont {Itoh}},  \emph
  {et~al.},\ }\href {https://doi.org/10.1038/s41467-021-24371-7} {\bibfield
  {journal} {\bibinfo  {journal} {Nat. Commun.}\ }\textbf {\bibinfo {volume}
  {12}},\ \bibinfo {pages} {4114} (\bibinfo {year} {2021})}\BibitemShut
  {NoStop}%
\bibitem [{\citenamefont {Gaudreau}\ \emph {et~al.}(2012)\citenamefont
  {Gaudreau}, \citenamefont {Granger}, \citenamefont {Kam}, \citenamefont
  {Aers}, \citenamefont {Studenikin}, \citenamefont {Zawadzki}, \citenamefont
  {Pioro-Ladriere}, \citenamefont {Wasilewski},\ and\ \citenamefont
  {Sachrajda}}]{gaudreau2012coherent}%
  \BibitemOpen
  \bibfield  {author} {\bibinfo {author} {\bibfnamefont {L.}~\bibnamefont
  {Gaudreau}}, \bibinfo {author} {\bibfnamefont {G.}~\bibnamefont {Granger}},
  \bibinfo {author} {\bibfnamefont {A.}~\bibnamefont {Kam}}, \bibinfo {author}
  {\bibfnamefont {G.}~\bibnamefont {Aers}}, \bibinfo {author} {\bibfnamefont
  {S.}~\bibnamefont {Studenikin}}, \bibinfo {author} {\bibfnamefont
  {P.}~\bibnamefont {Zawadzki}}, \bibinfo {author} {\bibfnamefont
  {M.}~\bibnamefont {Pioro-Ladriere}}, \bibinfo {author} {\bibfnamefont
  {Z.}~\bibnamefont {Wasilewski}}, \ and\ \bibinfo {author} {\bibfnamefont
  {A.}~\bibnamefont {Sachrajda}},\ }\href {https://doi.org/10.1038/nphys2149}
  {\bibfield  {journal} {\bibinfo  {journal} {Nat. Phys.}\ }\textbf {\bibinfo
  {volume} {8}},\ \bibinfo {pages} {54} (\bibinfo {year} {2012})}\BibitemShut
  {NoStop}%
\bibitem [{\citenamefont {Roszak}\ \emph {et~al.}(2015)\citenamefont {Roszak},
  \citenamefont {Filip},\ and\ \citenamefont
  {Novotn{\`y}}}]{roszak2015decoherence}%
  \BibitemOpen
  \bibfield  {author} {\bibinfo {author} {\bibfnamefont {K.}~\bibnamefont
  {Roszak}}, \bibinfo {author} {\bibfnamefont {R.}~\bibnamefont {Filip}}, \
  and\ \bibinfo {author} {\bibfnamefont {T.}~\bibnamefont {Novotn{\`y}}},\
  }\href {https://doi.org/10.1038/srep09796} {\bibfield  {journal} {\bibinfo
  {journal} {Sci. Rep.}\ }\textbf {\bibinfo {volume} {5}},\ \bibinfo {pages}
  {9796} (\bibinfo {year} {2015})}\BibitemShut {NoStop}%
\bibitem [{\citenamefont {Botzem}\ \emph {et~al.}(2016)\citenamefont {Botzem},
  \citenamefont {McNeil}, \citenamefont {Mol}, \citenamefont {Schuh},
  \citenamefont {Bougeard},\ and\ \citenamefont {Bluhm}}]{10.1038/ncomms11170}%
  \BibitemOpen
  \bibfield  {author} {\bibinfo {author} {\bibfnamefont {T.}~\bibnamefont
  {Botzem}}, \bibinfo {author} {\bibfnamefont {R.~P.~G.}\ \bibnamefont
  {McNeil}}, \bibinfo {author} {\bibfnamefont {J.-M.}\ \bibnamefont {Mol}},
  \bibinfo {author} {\bibfnamefont {D.}~\bibnamefont {Schuh}}, \bibinfo
  {author} {\bibfnamefont {D.}~\bibnamefont {Bougeard}}, \ and\ \bibinfo
  {author} {\bibfnamefont {H.}~\bibnamefont {Bluhm}},\ }\href
  {http://dx.doi.org/10.1038/ncomms11170} {\bibfield  {journal} {\bibinfo
  {journal} {Nat. Commun.}\ }\textbf {\bibinfo {volume} {7}},\ \bibinfo {pages}
  {11170} (\bibinfo {year} {2016})}\BibitemShut {NoStop}%
\bibitem [{\citenamefont {Neder}\ \emph {et~al.}(2007)\citenamefont {Neder},
  \citenamefont {Marquardt}, \citenamefont {Heiblum}, \citenamefont {Mahalu},\
  and\ \citenamefont {Umansky}}]{10.1038/nphys627}%
  \BibitemOpen
  \bibfield  {author} {\bibinfo {author} {\bibfnamefont {I.}~\bibnamefont
  {Neder}}, \bibinfo {author} {\bibfnamefont {F.}~\bibnamefont {Marquardt}},
  \bibinfo {author} {\bibfnamefont {M.}~\bibnamefont {Heiblum}}, \bibinfo
  {author} {\bibfnamefont {D.}~\bibnamefont {Mahalu}}, \ and\ \bibinfo {author}
  {\bibfnamefont {V.}~\bibnamefont {Umansky}},\ }\href {\doibase
  10.1038/nphys627} {\bibfield  {journal} {\bibinfo  {journal} {Nat. Phys.}\
  }\textbf {\bibinfo {volume} {3}},\ \bibinfo {pages} {534} (\bibinfo {year}
  {2007})}\BibitemShut {NoStop}%
\bibitem [{\citenamefont {Reilly}\ \emph {et~al.}(2008)\citenamefont {Reilly},
  \citenamefont {Taylor}, \citenamefont {Petta}, \citenamefont {Marcus},
  \citenamefont {Hanson},\ and\ \citenamefont
  {Gossard}}]{doi:10.1126/science.1159221}%
  \BibitemOpen
  \bibfield  {author} {\bibinfo {author} {\bibfnamefont {D.~J.}\ \bibnamefont
  {Reilly}}, \bibinfo {author} {\bibfnamefont {J.~M.}\ \bibnamefont {Taylor}},
  \bibinfo {author} {\bibfnamefont {J.~R.}\ \bibnamefont {Petta}}, \bibinfo
  {author} {\bibfnamefont {C.~M.}\ \bibnamefont {Marcus}}, \bibinfo {author}
  {\bibfnamefont {M.~P.}\ \bibnamefont {Hanson}}, \ and\ \bibinfo {author}
  {\bibfnamefont {A.~C.}\ \bibnamefont {Gossard}},\ }\href {\doibase
  10.1126/science.1159221} {\bibfield  {journal} {\bibinfo  {journal}
  {Science}\ }\textbf {\bibinfo {volume} {321}},\ \bibinfo {pages} {817}
  (\bibinfo {year} {2008})}\BibitemShut {NoStop}%
\bibitem [{\citenamefont {Mavadia}\ \emph {et~al.}(2017)\citenamefont
  {Mavadia}, \citenamefont {Frey}, \citenamefont {Sastrawan}, \citenamefont
  {Dona},\ and\ \citenamefont {Biercuk}}]{mavadia2017prediction}%
  \BibitemOpen
  \bibfield  {author} {\bibinfo {author} {\bibfnamefont {S.}~\bibnamefont
  {Mavadia}}, \bibinfo {author} {\bibfnamefont {V.}~\bibnamefont {Frey}},
  \bibinfo {author} {\bibfnamefont {J.}~\bibnamefont {Sastrawan}}, \bibinfo
  {author} {\bibfnamefont {S.}~\bibnamefont {Dona}}, \ and\ \bibinfo {author}
  {\bibfnamefont {M.~J.}\ \bibnamefont {Biercuk}},\ }\href
  {https://doi.org/10.1038/ncomms14106} {\bibfield  {journal} {\bibinfo
  {journal} {Nat. Commun.}\ }\textbf {\bibinfo {volume} {8}},\ \bibinfo {pages}
  {14106} (\bibinfo {year} {2017})}\BibitemShut {NoStop}%
\bibitem [{\citenamefont {Nakajima}\ \emph {et~al.}(2018)\citenamefont
  {Nakajima}, \citenamefont {Delbecq}, \citenamefont {Otsuka}, \citenamefont
  {Amaha}, \citenamefont {Yoneda}, \citenamefont {Noiri}, \citenamefont
  {Takeda}, \citenamefont {Allison}, \citenamefont {Ludwig}, \citenamefont
  {Wieck} \emph {et~al.}}]{nakajima2018coherent}%
  \BibitemOpen
  \bibfield  {author} {\bibinfo {author} {\bibfnamefont {T.}~\bibnamefont
  {Nakajima}}, \bibinfo {author} {\bibfnamefont {M.~R.}\ \bibnamefont
  {Delbecq}}, \bibinfo {author} {\bibfnamefont {T.}~\bibnamefont {Otsuka}},
  \bibinfo {author} {\bibfnamefont {S.}~\bibnamefont {Amaha}}, \bibinfo
  {author} {\bibfnamefont {J.}~\bibnamefont {Yoneda}}, \bibinfo {author}
  {\bibfnamefont {A.}~\bibnamefont {Noiri}}, \bibinfo {author} {\bibfnamefont
  {K.}~\bibnamefont {Takeda}}, \bibinfo {author} {\bibfnamefont
  {G.}~\bibnamefont {Allison}}, \bibinfo {author} {\bibfnamefont
  {A.}~\bibnamefont {Ludwig}}, \bibinfo {author} {\bibfnamefont {A.~D.}\
  \bibnamefont {Wieck}},  \emph {et~al.},\ }\href
  {https://doi.org/10.1038/s41467-018-04544-7} {\bibfield  {journal} {\bibinfo
  {journal} {Nat. Commun.}\ }\textbf {\bibinfo {volume} {9}},\ \bibinfo {pages}
  {2133} (\bibinfo {year} {2018})}\BibitemShut {NoStop}%
\bibitem [{\citenamefont {Calderon-Vargas}\ and\ \citenamefont
  {Kestner}(2015)}]{PhysRevB.91.035301}%
  \BibitemOpen
  \bibfield  {author} {\bibinfo {author} {\bibfnamefont {F.~A.}\ \bibnamefont
  {Calderon-Vargas}}\ and\ \bibinfo {author} {\bibfnamefont {J.~P.}\
  \bibnamefont {Kestner}},\ }\href {\doibase 10.1103/PhysRevB.91.035301}
  {\bibfield  {journal} {\bibinfo  {journal} {Phys. Rev. B}\ }\textbf {\bibinfo
  {volume} {91}},\ \bibinfo {pages} {035301} (\bibinfo {year}
  {2015})}\BibitemShut {NoStop}%
\end{thebibliography}

%

\end{document}